\documentclass[12pt]{article}
\usepackage{amsmath}
\usepackage{amssymb}
\usepackage{graphicx}
\usepackage{caption2}
\usepackage{amsfonts}
\usepackage{cite}
\usepackage{lineno}
\usepackage{color}
\usepackage{multirow}
\usepackage{subfigure}
\usepackage{bm}

\newcommand{\be}{\begin{equation}}
\newcommand{\ee}{\end{equation}}
\newcommand{\bea}{\begin{eqnarray}}
\newcommand{\eea}{\end{eqnarray}}
\newcommand{\bef}{\begin{figure}}
\newcommand{\ef}{\end{figure}}
\newcommand{\bt}{\begin{tabular}}
\newcommand{\et}{\end{tabular}}
\newcommand{\bno}{\begin{enumerate}}
\newcommand{\eno}{\end{enumerate}}

\def\3{\ss}

\catcode`\"=\active
\def"{\accent'177}

\begin{document}

\begin{center}
{\bf\large  Spatial symmetry invariance of three-dimensional Navier-Stokes turbulence}

Shijun Liao\footnote{Corresponding author, Email: sjliao@sjtu.edu.cn}\\

State Key Laboratory of Ocean Engineering, Shanghai, China\\
School of Ocean and Civil Engineering, Shanghai Jiao Tong University, China
\end{center}

\begin{abstract}

Direct Numerical Simulation (DNS) of the deterministic Navier-Stokes (NS) equations is widely used to provide benchmark solutions of turbulent flow, but its accuracy and reliability have not been rigorously inspected from a mathematical viewpoint. Here we prove a spatial symmetry invariance theorem (SSIT) of  three-dimensional NS equations whereby an exact solution of deterministic NS equations under periodic boundary conditions maintains the same spatial symmetry as its symmetric initial flow, forever.  SSIT provides an absolute benchmark test for computational fluid dynamics (CFD).  It is shown herein that traditional DNS of NS turbulence  quickly violates SSIT,  revealing that the spatiotemporal trajectory of NS turbulence predicted by DNS soon becomes badly polluted by numerical noise.  Importantly, it is found that even the statistical results of DNS might be unstable, i.e., exhibit sensitivity dependence on the initial flow conditions.  In other words, the loss of spatial symmetry in traditional DNS of NS turbulence is not a physical instability but instead a numerical artifact caused by the chaotic nature of the flow. Using the SSIT, it is straightforward to deduce several key properties of NS turbulence, including `ultra-chaos', the so-called `noise-expansion cascade', `permanent memory of symmetric initial flow conditions',  and so on.    Note that  permanent memory of symmetric initial flow conditions suggests that  NS turbulence governed by deterministic NS equations subject to symmetric initial flow should have no `mixing', corresponding to non-ergodicity, although ergodicity is one of the most fundamental assumptions of turbulence theory.  SSIT is therefore capable of greatly deepening our physical insight concerning NS turbulence.  In the future, SSIT could provide a completely new theoretical cornerstone and a completely novel perspective regarding turbulence,  considered by Feynman to be the most important unsolved problem of classical physics.  

\end{abstract}

\hspace{-0.4cm}{\bf Keyword} Navier-Stokes equations, spatial symmetry invariance  theorem, SSIT, turbulence, ergodicity, mixing

\hspace{-0.4cm}{\bf MSC} 76F02, 76F65, 65M25

\section{Introduction}

Nowadays direct numerical simulation (DNS)~\cite{Orszag1970, Rogallo1981NASA, She1990Nature, MoinARFM1998, Coleman2010DNS,FeracoScience2024} has been widely applied to solve Navier-Stokes (NS) turbulence,  i.e. turbulence governed by NS equations~\cite{Navier1822, Stokes1845}.  However, given that  exact solutions of NS turbulence are unknown, there exists a reliability crisis in turbulence simulation:  it is unknown whether numerical results are true solutions in mathematics or just  artifacts caused by numerical noise. 

Currently, many researchers~\cite{Deissler1986PoF, boffetta2001predictability, boffetta2017chaos, berera2018chaotic, Vassilicos2023JFM, Okamoto2025PRF} pointed out that NS turbulence is chaotic in that the spatiotemporal trajectories of NS turbulence are very sensitive to small disturbances.  Numerical algorithms had to be used because the exact solutions of chaotic systems are unknown.   But all numerical algorithms contain numerical noise, in the form of truncation and round-off errors.   In 2006 Lorenz~\cite{Lorenz2006Tellus} further discovered that numerical noise might exert distinct influences on the macroscopic properties of chaotic systems.  Invariably, the traditional algorithms of DNS for NS turbulence
introduce numerical error, which may increase exponentially to a macro-level in a chaotic flow. This suggests that the DNS result of NS turbulence should be quickly, badly polluted by numerical noise.  To confirm this logical deduction, Liao~\cite{Liao2009}  proposed the ``clean numerical simulation'' (CNS) in 2009.  Unlike DNS whose algorithms are solved using double precision (i.e., with a fixed level of round-off error), CNS~\cite{Liao2013, Liao2014SCPMA, Li2017SCPMA, Li2018PASJ, Liao2022NA, Hu2020JCP, Qin2020CSF, Liao2022AAMM, Liao2023book} reduced both the truncation error (by using high-order Taylor expansions in time and a Fourier spectral method in space) and the round-off error (by using multiple precision arithmetic) to such predetermined minimum level that numerical noise remained negligible (compared to the exact solution) throughout a finite time period that was sufficiently long to accurately compute the flow statistics.    Using CNS as a new simulation tool to gain clean turbulence, Qin and Liao \cite{Qin2022JFM} found that numerical noise as a stochastic disturbance could indeed have significant impact on sustained turbulence, not only in terms of its spatiotemporal trajectory but also in its statistics and macro-level properties (such as flow type).  Notably, for NS turbulence subject to a symmetric initial condition, all CNS results maintain the same spatial symmetry as for the initial conditions in a finite but long enough interval of time.   However, all corresponding DNS results (without enforcing the same spatial symmetry for $t > 0$) maintain the same spatial symmetry as for the initial conditions only over a short initial period after which symmetry is completely lost~\cite{Qin2024JOES, Liao-2025-JFM-NEC, Qin2025PoF, Qin2026EJMB}, indicating that numerical noise, which is random in space, indeed increases to a macro-level and then destroys the spatial symmetry of DNS results.
                
In 1985, Kida~\cite{Kida1985} used numerical simulation to investigate a case of two-dimensional turbulence.   According to the covariation of NS equations, Kida~\cite{Kida1985} assumed that the turbulent flow should maintain the same spatial symmetry as its initial condition.   To increase computational efficiency, Kida used spatial symmetry to reduce the size of the computational domain by a factor of 64 so as to achieve higher Reynolds number flows with limited resources. This approach treated spatial symmetry as an observed utility rather than a proven law~\cite{Kida1985}.  However, in violation of this assumption,  almost all current state-of-the-art DNS (without using the property of spatial symmetry to increase computational efficiency) of NS turbulence subject to a symmetric initial condition  maintains the same spatial symmetry as the  initial condition only over a short period of time after the simulation commences, after which the flow then rapidly loses its spatial symmetry~\cite{Qin2024JOES, Liao-2025-JFM-NEC, Qin2025PoF, Qin2026EJMB}.    
Note that DNS results have been widely used as benchmark solutions by the turbulence community for more than a half century~\cite{Orszag1970, Rogallo1981NASA, She1990Nature, MoinARFM1998, Coleman2010DNS, FeracoScience2024}.   Note that all current state-of-the-art DNS results (obtained without using spatial symmetry to increase computational efficiency) of NS turbulence subject to an initial condition with spatial symmetry rapidly violate Kida's assumption of fully symmetric flow~\cite{Qin2024JOES, Liao-2025-JFM-NEC, Qin2025PoF, Qin2026EJMB}.   Should we trust the DNS results in such cases? 
Besides, it has been established that CNS results maintain the same spatial symmetry as the initial condition over a prescribed long, finite period~\cite{Qin2024JOES, Liao-2025-JFM-NEC, Qin2025PoF, Qin2026EJMB}.  It is therefore unknown whether or not the exact solution of NS turbulence subject to a symmetric initial condition should maintain the same spatial symmetry forever. Note that, despite the widespread use of spatial symmetry, a global mathematical proof of symmetry persistence is a missing `cornerstone' of fluid mechanics.   This implies the existence of a  `reliability crisis' in turbulence simulation, as mentioned earlier.

Physical insights concerning turbulent flows can be obtained by physical experiments involving actual turbulence,  numerical simulation  and  mathematical investigation of a  mathematical model of turbulence.  Although numerical simulation is routinely used nowadays, it is very beneficial to have access to theoretical results based on purely mathematical proofs of NS turbulence without recourse to numerical simulations. 
 In \S~2,  a mathematical proof is given of the spatial symmetry invariance theorem (SSIT), whereby the exact solution of the NS equations under periodic boundary conditions maintains the same spatial symmetry as its symmetric initial flow, forever.  Numerical confirmation and applications of the SSIT are described in \S~3 which illustrates that certain important properties of NS turbulence, including `ultra-chaos' and the `noise-expansion cascade' can be straightforwardly deduced from the SSIT.  It is demonstrated that the spatiotemporal trajectory of traditional DNS is quickly badly polluted by numerical noise, as the DNS results rapidly violate the SSIT.    It is found that the loss of symmetry in traditional DNS is not a `physical instability' but instead a `numerical artifact' caused by numerical nose and the chaotic nature of NS turbulence. The foregoing enables deeper physical insight into turbulence. \S~4  presents a detailed discussion of the findings.  \S~5 draws the main conclusions.
 
\section{Spatial symmetry invariance of NS turbulence}

Consider a dimensionless three-dimensional (3D) incompressible NS equations~\cite{Batchelor2000}: 
\begin{eqnarray}
\nabla \cdot {\bf u} &=& 0,  \label{eq:continuation} \\
 \frac{\partial {\bf u}}{\partial t}
 +\left( {\bf u}\cdot \nabla\right) {\bf u} &=&-\nabla p + \frac{1}{Re} \Delta {\bf u} + {\bf f},       \label{eq:u}
\end{eqnarray}
subject to periodic boundary conditions  in a square domain ${\bf x}=(x_1,x_2,x_3)\in[0,2\pi]^3$,   
where $t\in[0,+\infty)$ denotes time, ${\bf u}=(u_1,u_2,u_3)$ is the fluid flow velocity,  $p$ is pressure, ${\bf f}({\bf x})$ is a steady-state external force,  $Re$ is the Reynolds number based on a characteristic length,  $\nabla$ is a Hamilton operator,  and $\Delta=\nabla\cdot \nabla$ is a Laplace operator,  respectively.  Let $O-x_{1}x_{2}x_{3}$ denote its corresponding coordinate system.

Define 
\begin{equation}
 {\bf u}^{(m)}({\bf x},t) =\frac{1}{m!} \frac{\partial {\bf u}({\bf x},t)}{\partial t^{m}},\;\;\;  p^{(m)}({\bf x},t) =\frac{1}{m!}  \frac{\partial p({\bf x},t)}{\partial t^{m}}, \;\;\; m\geq 0, \label{def:u[m]:p[m]}
 \end{equation}
 with 
 \begin{equation}
  {\bf u}^{(0)}({\bf x},t) =  {\bf u}({\bf x},t), \;\;\;  p^{(0)}({\bf x},t) = p({\bf x},t),
 \end{equation}
for integer $m\geq 0$.  

From (\ref{eq:u}), it holds that 
\begin{eqnarray}
 \frac{\partial (\nabla\cdot {\bf u})}{\partial t}
 +\nabla\cdot  \left[\left( {\bf u}\cdot \nabla\right) {\bf u}\right] &=&-\Delta p + \frac{1}{Re} \Delta \left(\nabla\cdot {\bf u}\right) + \nabla\cdot {\bf f},    
\end{eqnarray}
which,  using the continuation equation (\ref{eq:continuation}), gives  that
\begin{equation}
\Delta p =\nabla\cdot {\bf f} -\nabla\cdot  \left[\left( {\bf u}\cdot \nabla\right) {\bf u}\right]. \label{eq:p}    
\end{equation}
Differentiating (\ref{eq:u}) and (\ref{eq:p}) $m$ times with respect to $t$ and then dividing by $m!$, we have
\begin{equation}
(m+1){\bf u}^{(m+1)} 
 +\sum_{i=0}^{m}\left( {\bf u}^{(i)}\cdot \nabla\right) {\bf u}^{(m-i)} =-\nabla p^{(m)} + \frac{1}{Re} \Delta {\bf u}^{(m)} ,       \;\;  \label{eq:u:m}
\end{equation}
\begin{equation}
\Delta p^{(m)} =-\sum_{i=0}^{m}\nabla\cdot  \left[\left( {\bf u}^{(i)}\cdot \nabla\right) {\bf u}^{(m-i)}\right], \label{eq:p:m}    
\end{equation}
for $m\geq 1$ and $t\geq 0$.

Let ${\bf u}'({\bf x}',t)$, $p'({\bf x}',t)$, ${\bf f}'({\bf x}')$ denote the velocity and pressure of fluid, and the external force  in the coordinate system $O'-x'_{1}x'_{2}x'_{3}$ , where ${\bf x}'\in[0,2\pi]^3$.  Given that  they  also satisfy (\ref{eq:u}) and (\ref{eq:p})-(\ref{eq:p:m}) for $t\geq 0$, then 
\begin{eqnarray}
{\bf u}'^{(1)}
 +\left( {\bf u}'\cdot \nabla'\right) {\bf u}' &=&-\nabla' p' + \frac{1}{Re} \Delta' {\bf u}' + {\bf f}',  \label{eq:u:B} \\
(m+1){\bf u}'^{(m+1)} 
 +\sum_{i=0}^{m}\left( {\bf u}'^{(i)}\cdot \nabla'\right) {\bf u'^{(m-i)}} &=& -\nabla' p'^{(m)} + \frac{1}{Re} \Delta' {\bf u}'^{(m)} ,   \;\; m\geq 1,     \label{eq:u:B:m}
\end{eqnarray}
and 
\begin{eqnarray}
\Delta' p' &=& \nabla'\cdot {\bf f}' -\nabla'\cdot  \left[\left( {\bf u'}\cdot \nabla'\right) {\bf u'}\right],  \label{eq:p:B}   \\ 
\Delta' p'^{(m)} &=&-\sum_{i=0}^{m}\nabla'\cdot  \left[\left( {\bf u}'^{(i)}\cdot \nabla'\right) {\bf u}'^{(m-i)}\right], \;\; m\geq 1. \label{eq:p:m:B}    
\end{eqnarray}

In general,  ${\bf u}'({\bf x}',t) \neq {\bf u}({\bf x},t)$.   However, if there exists ${\bf u}'({\bf x}',t) = {\bf u}({\bf x},t)$, then there also exists spatial symmetry under the coordinate translation $x'_{i}=\alpha_i+x_{i}$ and/or the coordinate rotation $x'_{i}=\beta_i-x_{i}$ with constant  $\alpha_i$ and $\beta_i$ $(i=1,2,3)$.  

\begin{description}
\item[Lemma A (Operator Covariance)]  {\em Under the coordinate translation $x'_{i}=x_{i}+\alpha_i$ and/or the coordinate rotation $x'_{i}=\beta_i-x_{i}$, where $\alpha_i$ and $\beta_i$ $(i=1,2,3)$ are constant,  it holds that}
\begin{equation}
\nabla' = \nabla, \;\; \Delta' = \Delta.   \label{Lemma:A}
\end{equation}  
\end{description}

\hspace{-0.75cm}Proof:  Let ${\bf e}_{i}, {\bf e}'_{i}$ denote axis unit vectors for $x_{i}, x'_{i}$, where $i=1,2,3$.    For the coordinate translation $x'_{i}=x_{i}+\alpha_i$ with constant $\alpha_i$ $(i=1,2,3)$,  it holds  that 
\begin{equation}
{\bf e}_{i} = {\bf e}'_{i}, \;\;\; \frac{\partial }{\partial x_{i}} = \frac{\partial }{\partial x'_{i}} \frac{\partial x'_{i}}{\partial x_{i}} = \frac{\partial }{\partial x'_{i}}, \;\;\; i=1,2,3,
\end{equation}  
and so
\begin{equation}
\nabla' =\sum_{i=1}^{3} {\bf e}'_{i} \;  \frac{\partial }{\partial x'_{i}} =\sum_{i=1}^{3} {\bf e}_{i} \;  \frac{\partial }{\partial x_{i}} = \nabla.
\end{equation}
Similarly, for the coordinate rotation $x'_{i}=\beta_i-x_{i}$, with constant $\beta_i$ $(i=1,2,3)$, it holds that  
\begin{equation}
{\bf e}_{i} = -{\bf e}'_{i}, \;\;\; \frac{\partial }{\partial x_{i}} = \frac{\partial }{\partial x'_{i}} \frac{\partial x'_{i}}{\partial x_{i}} = -\frac{\partial }{\partial x'_{i}},
\end{equation}  
and so  
\begin{equation}
\nabla' =\sum_{i=1}^{3} {\bf e}'_{i} \;  \frac{\partial }{\partial x'_{i}} =\sum_{i=1}^{3} \left(-{\bf e}_{i}\right) \; \left(- \frac{\partial }{\partial x_{i}}\right) =\sum_{i=1}^{3} {\bf e}_{i} \;  \frac{\partial }{\partial x_{i}} = \nabla.
\end{equation}
So, for  $x'_{i}=x_{i}+\alpha_i$ and/or   $x'_{i}=\beta_i-x_{i}$ $(i=1,2,3)$, we have 
\begin{equation}
\Delta' = \nabla' \cdot \nabla' = \nabla \cdot \nabla =\Delta. 
\end{equation}
This ends the proof of Lemma A whereby the Hamilton and Laplacian operators are covariant under specific spatial transformations.

\begin{description}
\item[Lemma B (Recursive Symmetry)]  {\em 
For the incompressible Navier-Stokes equations (\ref{eq:continuation}) and (\ref{eq:u}) subject to  periodic boundary conditions, if there exist the spatial symmetry ${\bf u}'({\bf x}',t_{0}) = {\bf u}({\bf x},t_{0})$ at $t=t_{0}$ and ${\bf f}'({\bf x}')= {\bf f}({\bf x})$, where  $x'_i = \alpha_i+x_i$ and/or  $x'_i = \beta_i-x_i$ with constant $\alpha_i$ and $\beta_i$ $(i=1,2,3)$, and besides  there exists a temporal Taylor series  of ${\bf u}({\bf x},t)$ at $t=t_{0}$, then there exist spatial symmetry  ${\bf u}'^{(m+1)}({\bf x}',t_{0}) = {\bf u}^{(m+1)}({\bf x},t_{0})$ and   $p'^{(m)}({\bf x}',t_{0}) = p^{(m)}({\bf x},t_{0})$ at $t=t_{0}$ for all integer $m\geq 0$. 
}
\end{description}

\hspace{-0.75cm}Proof.  Lemma~B can be proved by means of recursive method. 

\hspace{-0.75cm} (A) When $m=0$, (\ref{eq:p:B}) becomes at $t=t_{0}$ that 
\begin{equation}
\Delta p'=\nabla\cdot {\bf f}  -\nabla\cdot  \left[\left( {\bf u}\cdot \nabla\right) {\bf u}\right], \label{eq:p:B:2}    
\end{equation}
given that  $\Delta' =\Delta, \nabla' = \nabla$ according to Lemma~A, and besides there exist the spatial symmetry ${\bf u}'({\bf x}',t_0) = {\bf u}({\bf x},t_{0})$ and ${\bf f}'({\bf x}') = {\bf f}({\bf x})$.   Comparing (\ref{eq:p:B:2}) with (\ref{eq:p}), we have  $\Delta p =\Delta p'$, such that 
$p'({\bf x}',t_{0}) = p({\bf x},t_{0})$, i.e.  $p'^{(0)}({\bf x}',t_{0}) = p^{(0)}({\bf x},t_{0})$.

Similarly, (\ref{eq:u:B}) becomes at $t=t_{0}$ that
 \begin{equation}
 {\bf u}'^{(1)}
 +\left( {\bf u}\cdot \nabla\right) {\bf u} =-\nabla p + \frac{1}{Re} \Delta {\bf u} + {\bf f},  \label{eq:u:B:2}
 \end{equation}
because $\Delta' =\Delta, \nabla' = \nabla$ according to Lemma~A and there exist the spatial symmetries  ${\bf u}'({\bf x}',t_0) = {\bf u}({\bf x},t_{0})$, ${\bf f}'({\bf x}') = {\bf f}({\bf x})$ and $p'({\bf x}',t_{0}) = p({\bf x},t_{0})$ as mentioned  above.  Comparing  (\ref{eq:u:B:2}) to (\ref{eq:u}) gives ${\bf u}'^{(1)}({\bf x}',t_0) = {\bf u}^{(1)}({\bf x},t_{0})$.  
 
 Thus, Lemma~B holds when $m=0$.  

\hspace{-0.75cm}(B) Assume that Lemma~B holds when $m = n \geq 0$,   and  there exists the spatial symmetry 
 \begin{equation}
 {\bf u}'^{(i+1)}({\bf x}',t_{0}) = {\bf u}^{(i+1)}({\bf x},t_{0}),  \;\;  p'^{(i)}({\bf x}',t_{0}) = p^{(i)}({\bf x},t_{0}), \;\; 0\leq i\leq n. \label{symmetry:n}
 \end{equation} 
  Then, (\ref{eq:p:m:B} ) becomes  at $t=t_{0}$ that 
 \begin{equation}
 \Delta p'^{(n+1)} =-\sum_{i=0}^{n+1}\nabla\cdot  \left[\left( {\bf u}^{(i)}\cdot \nabla\right) {\bf u}^{(n+1-i)}\right],  \label{eq:p:n}
 \end{equation}
because $\Delta' =\Delta, \nabla' = \nabla$ according to Lemma~A and there exist the spatial symmetries  given by (\ref{symmetry:n}).  Comparing (\ref{eq:p:n}) to (\ref{eq:p:m}) gives  $p'^{(n+1)}({\bf x}',t_{0}) = p^{(n+1)}({\bf x},t_{0})$.   
 
 Then, according to the spatial symmetry (\ref{symmetry:n}), (\ref{eq:u:B:m}) (when $m=n+1$) becomes at $t=t_{0}$  that  
  \begin{equation}
 (n+2){\bf u}'^{(n+2)} 
 +\sum_{i=0}^{n+1}\left( {\bf u}^{(i)}\cdot \nabla\right) {\bf u}^{(n+1-i)} = -\nabla p^{(n+1)} + \frac{1}{Re} \Delta {\bf u}^{(n+1)},   \label{eq:u:n:B}
 \end{equation}
because $\Delta' =\Delta, \nabla' = \nabla$ according to Lemma~A and  $p'^{(n+1)}({\bf x}',t_{0}) = p^{(n+1)}({\bf x},t_{0})$ as  proved above.   Setting $m=n+1$ in (\ref{eq:u:m}) gives at $t=t_{0}$ that
\begin{equation}
(n+2){\bf u}^{(n+2)} 
 +\sum_{i=0}^{n+1}\left( {\bf u}^{(i)}\cdot \nabla\right) {\bf u}^{(n+1-i)} =-\nabla p^{(n+1)} + \frac{1}{Re} \Delta {\bf u}^{(n+1)}.  \;\;  \label{eq:u:n}
\end{equation}
Comparing (\ref{eq:u:n:B}) to (\ref{eq:u:n}) gives ${\bf u}'^{(n+2)}({\bf x}',t_{0}) ={\bf u}^{(n+2)}({\bf x},t_{0})$.  

Thus, if Lemma~B holds for $m=n\geq 0$, it holds for $m=n+1$.

 \hspace{-0.75cm}(C) According to (A) and (B), Lemma~B holds for arbitrary integer $m\geq 0$.\\
 This ends the proof of Lemma~B.  It has utilized mathematical induction to show that if the initial condition is symmetric, all higher-order temporal derivatives must inherit exactly the same symmetry.  
  
 \begin{description}
\item[ Lemma C (Symmetry Continuation)]  {\em For the incompressible continuity and Navier-Stokes equations (\ref{eq:continuation}) and (\ref{eq:u}) subject to periodic boundary conditions, if there exist the spatial symmetries ${\bf u}'({\bf x}',t_{0}) = {\bf u}({\bf x},t_{0})$ at $t=t_{0}$ and ${\bf f}'({\bf x}')= {\bf f}({\bf x})$, where  $x'_i = \alpha_i+x_i$ and/or  $x'_i = \beta_i-x_i$ with constant $\alpha_i$ and $\beta_i$ $(i=1,2,3)$,   and besides  there exists  a temporal Taylor series  for ${\bf u}({\bf x},t)$ at $t=t_{0}$, with a non-zero radius $\rho$ of convergence, then there exists the spatial symmetry  ${\bf u}'({\bf x}',t) = {\bf u}({\bf x},t)$ in $t\in[t_{0},\rho)$. }
\end{description}

\hspace{-0.75cm}Proof.  According to Lemma~B, the temporal Taylor series of ${\bf u}({\bf x},t)$ at $t=t_{0}$ reads
\begin{eqnarray}
{\bf u}({\bf x},t) &=& \sum_{m=0}^{+\infty}  {\bf u}^{(m)}({\bf x},t_{0}) (t-t_{0})^{m}= \sum_{m=0}^{+\infty}  {\bf u}'^{(m)}({\bf x}',t_{0}) (t-t_{0})^{m}={\bf u}'({\bf x}',t) 
\end{eqnarray}
in $|t-t_{0}|<\rho_{0}$, which includes $t\in[t_{0},t_{0}+\rho_{0})$.  This ends the proof of Lemma~C.

Finally, let us give the mathematical proof of the following theorem:
\begin{description}
\item[Spatial Symmetry Invariance Theorem (SSIT)] {\em 
The exact solution $\bf u$  of  incompressible  three-dimensional continuity and  Navier-Stokes equations (\ref{eq:continuation}) and (\ref{eq:u}) subject to periodic boundary conditions have the same spatial symmetry ${\bf u}'({\bf x}',t)$ $ =  {\bf u}({\bf x},t)$ for all $t>0$ as its smooth initial condition, where $x'_i = \alpha_{i}+x_i$   and/or   $x'_i = \beta_{i}-x_i$  with constants  $\alpha_{i}$ and $\beta_{i}$ $(i=1,2,3)$, provided there exist spatial symmetries ${\bf u}'({\bf x}',0)= {\bf u}({\bf x},0)$ for initial condition and  ${\bf f}'({\bf x}')= {\bf f}({\bf x})$ for the external force, and besides a temporal Taylor series for ${\bf u}({\bf x},t)$ exists with non-zero radius of convergence for $t\geq 0$. 
}
\end{description}

\hspace{-0.75cm}Proof.  Since there exists the spatial symmetries  ${\bf u}'({\bf x}',0) = {\bf u}({\bf x},0)$ and ${\bf f}'({\bf x}')= {\bf f}({\bf x})$, where $x'_i = \alpha_i + x_i$ and/or  $x'_i = \beta_i - x_i$  with  constant $\alpha_i$ and $\beta_i$ $(i=1,2,3$),  we have according to Lemma~C  (setting $t_{0}=0$) the spatial symmetry
\[  {\bf u}'({\bf x}',t) = {\bf u}({\bf x},t), \hspace{1.0cm} t\in[0,\delta_{0}],  \]
where $\delta_{0} <\rho_{0}$, and $\rho_{0}>0$ is the non-zero radius of convergence of its Taylor series at $t=0$.  

Similarly, since  there exists the spatial symmetries  ${\bf u}'({\bf x}',\delta_{0}) = {\bf u}({\bf x},\delta_{0})$ at $t=\delta_{0}$ and ${\bf f}'({\bf x}')= {\bf f}({\bf x})$, we have according to Lemma~C  (setting $t_{0}=\delta_{0}$) the spatial symmetry
\[  {\bf u}'({\bf x}',t) = {\bf u}({\bf x},t), \hspace{1.0cm} t\in[\delta_{0},\delta_{1}],  \]
where $\delta_{1} <\rho_{1}$, and $\rho_{1}>0$ is the non-zero radius of convergence of its Taylor series at $t=\delta_{0}$.

Note that 
\[   [0,\delta_{0}] \cup [\delta_{0},\delta_{1}] = [0,\delta_{1}].  \]
Hence,  there exists the spatial symmetry 
\[ {\bf u}'({\bf x}',t) = {\bf u}({\bf x},t), \hspace{1.0cm} t\in[0,\delta_{1}],  \]
over a larger interval of time.  The same process can continue and the time interval becomes larger and larger because the temporal Taylor series of ${\bf u}({\bf x},t)$ invariably has a non-zero radius of convergence.  
Thus,  it holds the spatial symmetry 
\[  {\bf u}'({\bf x}',t) = {\bf u}({\bf x},t), \hspace{1.0cm} t\in[0,+\infty). \] 
This ends the proof of the spatial symmetry invariance theorem (SSIT).

Note that, for the two-dimensional Kolmogorov flow in the form of stream function,  Liao~\cite{Liao2026SymmetryInvariance, Liao2026TMP}  proved the similar theorem.  The mathematical proof  described  above for the three-dimensional deterministic NS equations indicates that the SSIT has the general meanings.   

\section{Numerical confirmation and application of SSIT}

It should be emphasized that the spatial symmetry invariance  theorem (SSIT) provides  an `absolute scale' for Computational Fluid Dynamics (CFD).  In other words,  if a DNS fails to maintain the same symmetry of a symmetric initial state, the numerical simulation is mathematically invalidated as a true solution of NS equations.

Assume that exact solution ${\cal F}({\bf x},t)$ of the continuity and NS equations has a spatial symmetry ${\cal F}({\bf x},t)={\cal F}({\bf x}',t)$ in ${\bf x}\in[a,b]^{3}$ for $t\in[0,+\infty]$.  In any numerical simulation, there should invariably exist some difference between ${\cal F}({\bf x},t)$ and ${\cal F}({\bf x}',t)$.  So, we define spatial symmetry loss (denoted by $\delta$) of a numerical simulation by
\begin{equation}
\delta(t) =\left\{ \frac{1}{(b-a)^{3}}\int_{a}^{b}\int_{a}^{b}\int_{a}^{b} \left[  {\cal F}({\bf x},t) - {\cal F}({\bf x}',t) \right]^{2}dx_{1} d x_{2} d x_{3}\right\}^{1/2}.
\end{equation} 
Obviously, spatial symmetry is maintained when $\delta \ll 1$, but is destroyed  when $\delta \sim O(1)$.   
 
\subsection{Two-dimensional Kolmogorov flow}

\begin{figure}
    \begin{center}
        \begin{tabular}{cc}
             \includegraphics[width=2.5in]{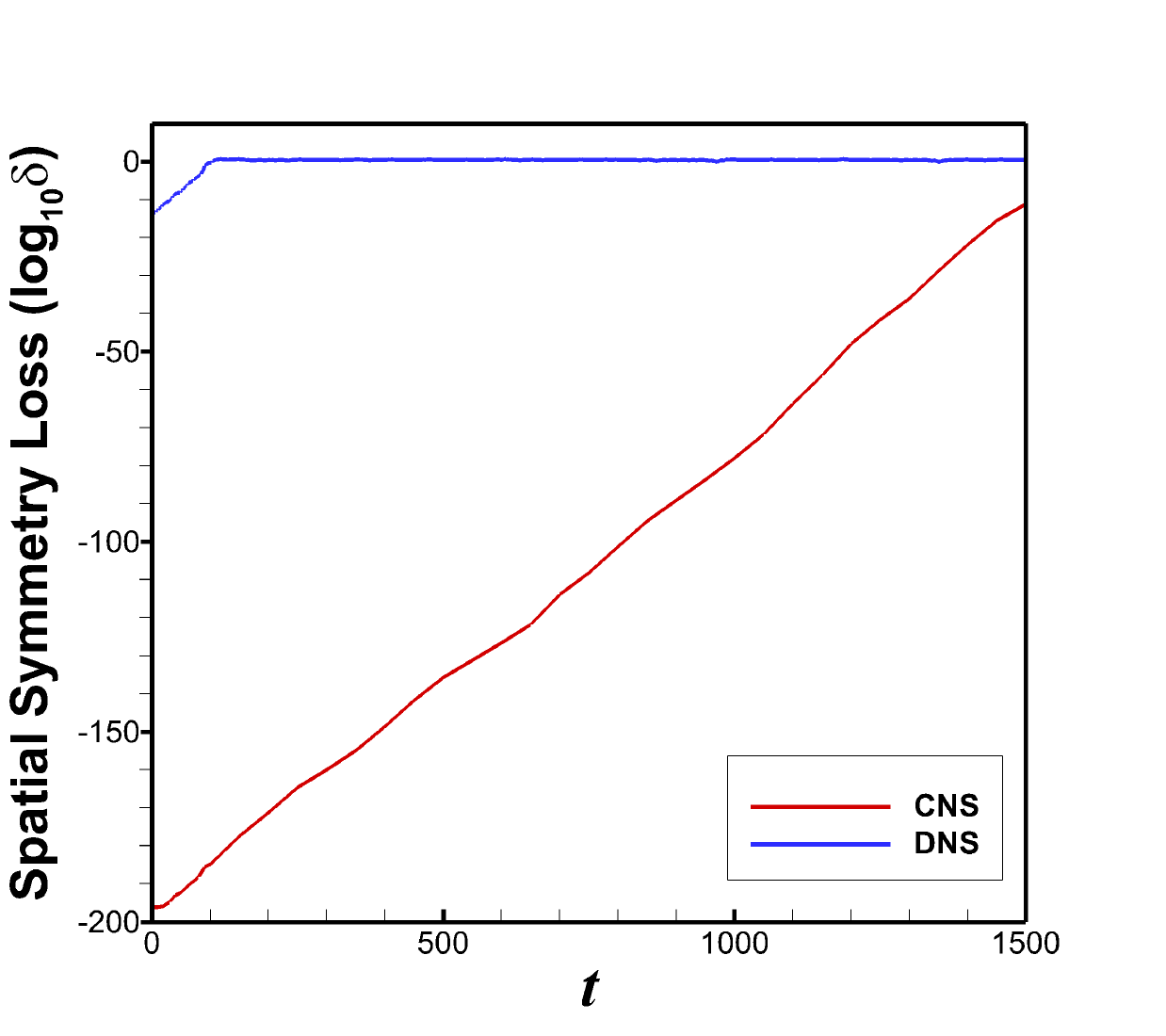}
             \includegraphics[width=2.5in]{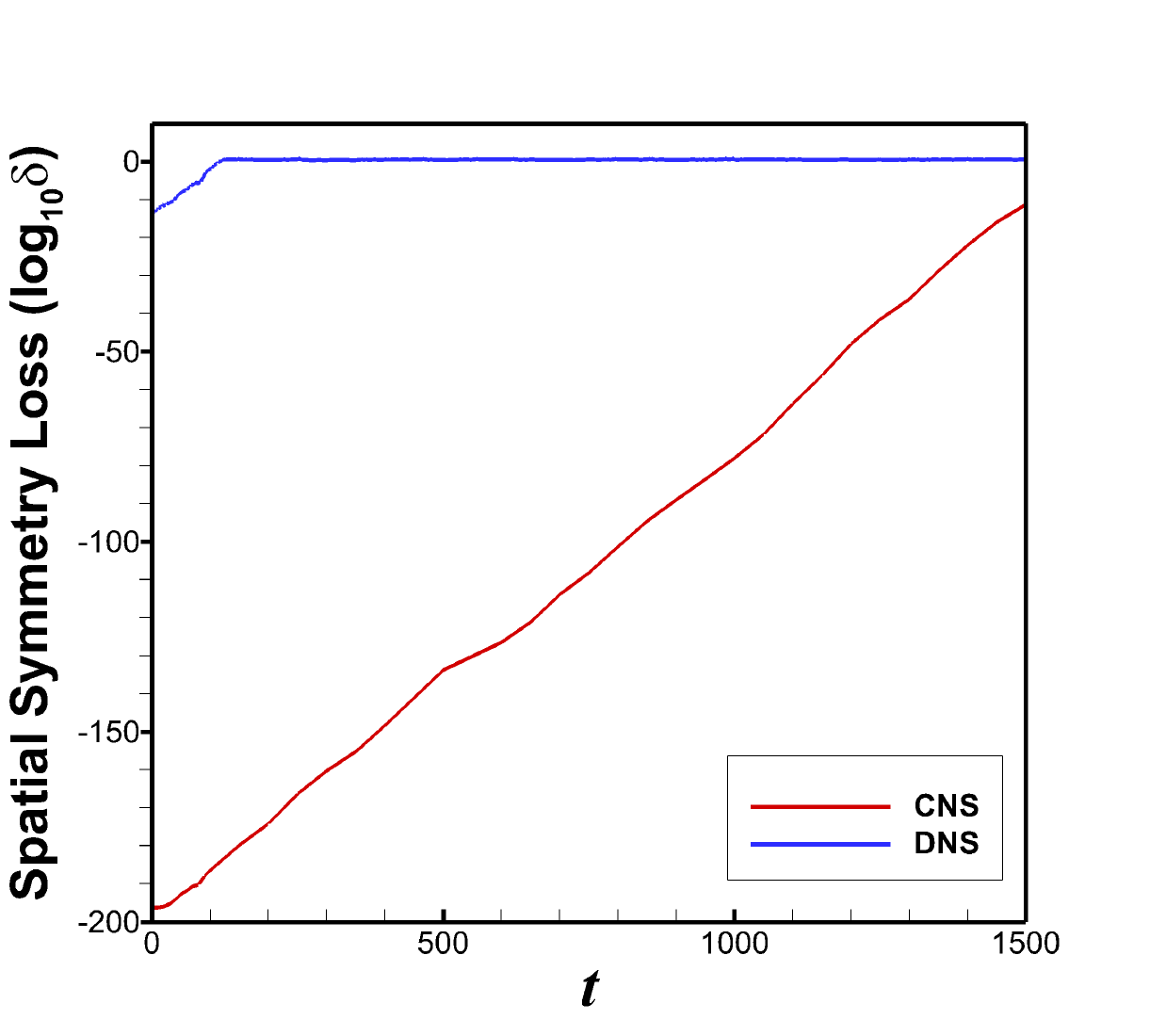}
        \end{tabular}
    \caption{Time histories of spatial symmetry loss given by CNS (red) and DNS (blue) for D Kolmogorov flow \cite{Qin2024JOES}.  Left:  spatial symmetry of translation ($s_{1}$); Right: spatial symmetry of rotation ($s_{2}$).  Spatial symmetry is defined by (\ref{symmetry-exam1}).}     \label{JOES}
    \end{center}
\end{figure}

Qin {\em et al.} \cite{Qin2024JOES} applied CNS to solve a two-dimensional (2D) Kolmogorov flow in the form of a stream-function $\psi$:
\begin{equation}
\frac{\partial}{\partial t} \left( \nabla^{2}\psi\right)+\frac{\partial(\psi,\nabla^{2}\psi)}{\partial(x,y)}-\frac{1}{Re}\nabla^{4}\psi+n_K\cos(n_Ky)=0,       \label{eq-Kolmogorov-2D}
\end{equation}
where $x,y\in[0,2\pi]$ are horizontal and vertical coordinates, $t$ denotes time,   $Re$  the Reynolds number, $\nabla^{2}$  a Laplace operator,  $\nabla^{4}=\nabla^{2}\nabla^{2}$, and 
\[
 \frac{\partial(a,b)}{\partial(x,y)}=\frac{\partial a}{\partial x}\frac{\partial b}{\partial y}-\frac{\partial b}{\partial x}\frac{\partial a}{\partial y}       \label{Jacobi}
\]
is the Jacobi operator, subject to the periodic  boundary condition 
\begin{equation}
\psi (x,y,t) = \psi (x\pm2\pi,y\pm 2\pi,t).  \label{psi-bc-periodic}
\end{equation}
Following Chandler and Kerswell \cite{chandler2013invariant}, Qin {\em et al.} \cite{Qin2024JOES}  prescribed  the physical parameters
$n_K$=4, $Re$=40 and the symmetric initial condition
\begin{align}
& \psi(x,y,0)=-\frac{1}{2}\big[ \cos(x+y)+\cos(x-y)\big],       \label{initial-condition-exam1}
\end{align}
corresponding to a sustained turbulent Kolmogorov flow.  

The initial condition (\ref{initial-condition-exam1}) has the spatial symmetries in translation ($s_{1}$) and rotation ($s_{2}$) given by 
\begin{equation}
\left\{
\begin{array} {lll}
\mbox{translation} &:& s_{1} \circ \psi(x,y,t) \mapsto \psi(x+\pi,y+\pi,t),\\
 \mbox{rotation} &:& s_{2} \circ \psi(x,y,t)  \mapsto  \psi(-x,-y,t), 
\end{array}
\right.  \label{symmetry-exam1}
\end{equation} 
respectively.  For  $n_k=4$,  Eq.~(\ref{eq-Kolmogorov-2D}) is covariant under the above spatial symmetries in translation and rotation, so that, according to the spatial symmetry invariance theorem (SSIT), its exact solution must maintain the same spatial symmetries (\ref{symmetry-exam1}) {\em forever}, i.e. for $t\in[0,+\infty)$.  

Qin {\em et al.} \cite{Qin2024JOES} used both of CNS and DNS to solve this 2D Kolmogorov flow.  Unlike Kida~\cite{Kida1985}, they did not use the spatial symmetry  (\ref{symmetry-exam1}) to increase computational efficiency of DNS and CNS.  In other words, their DNS and CNS algorithms are valid for a general initial condition, no matter whether it is symmetric or not.  
As shown in Fig.~\ref{JOES}, the spatial symmetry losses of the CNS result are much less than 1 over the whole simulation duration $t\in[0,1500]$, 
indicating that the CNS results maintain spatial symmetry for  in the whole simulation duration    $t\in[0,1500]$.  However,  as shown in Fig.~\ref{JOES},  the spatial symmetry losses of the DNS result increase to a macro-level $O(1)$ at $t\approx 110$,   indicating that the DNS result maintains the spatial symmetries only in a short duration $t\in[0,110]$,  after which time the results rapidly lose spatial symmetry.
Hence,  the DNS result for $t>110$ violates the SSIT. 
In other words,  based on the SSIT, we are quite certain that the DNS result after $t>110$ departs far away from the exact solution for turbulent Kolmogorov flow.  This test also clearly indicates that  loss of symmetry in traditional DNS is not due to a `physical instability'  but is in fact a `numerical artifact' caused by the chaotic nature of NS turbulence.  Moreover, Qin et al.~\cite{Qin2024JOES}  found that the statistics given by the CNS are distinctly different from those given by the DNS.   This is quite reasonable from the viewpoint of SSIT because the CNS results maintain spatial symmetry (\ref{symmetry-exam1})   but the DNS results rapidly lose symmetry due to stochastic numerical noise (that has no spatial symmetry at all) and the chaotic property of NS turbulence.

\subsection{Three-dimensional Kolmogorov flow}

Qin and Liao~\cite{Qin2025PoF} considered 3D Kolmogorov flow \cite{wu2021quadratic},  governed by the 3D incompressible Navier-Stokes equation
\begin{align}
\frac{\partial \textbf{u}}{\partial t}+(\textbf{u}\cdot\nabla)\textbf{u} = -\nabla p + \frac{1}{Re}\Delta\textbf{u} + \textbf{f},       \label{NS}
\end{align}
and the  continuity equation $\nabla\cdot\textbf{u}=0$,  where $t$ denotes time, $\textbf{x}=(x, y, z)\in[0,1]^3$ is the Cartesian coordinate, $\textbf{u}= u\; \textbf{e}_x +v\; \textbf{e}_y + w\; \textbf{e}_z$ is the velocity field,  $Re=1211.5$ is the Reynolds number, and the external force $\textbf{f}$ is of the form $\textbf{f} = \sin(8\pi  z)\,\textbf{e}_x$,  subject to a periodic boundary condition and the symmetric initial condition
\begin{equation}
\left\{
\begin{array}{l}
u(x,y,z,0)=\cos(2\pi x)\cos(2\pi y)\sin(2\pi z)/2,    \\
v(x,y,z,0)=\sin(2\pi x)\sin(2\pi y) \sin(2\pi z),  \\
w(x,y,z,0)=\sin(2\pi x)\cos(2\pi y)\cos(2\pi z)/2,   \\
\end{array}
\right.  \label{initial-condition-exam3}
\end{equation}
which has the spatial  symmetry of rotation ($s_{3}$)  
\begin{equation}
s_{3}\circ \textbf{u}(x,y,z,t)\mapsto -\textbf{u}(1-x,1-y,1-z,t).
\label{spatial-symmetry-exam3}
\end{equation} 
Note that the external force $\textbf{f} = \sin(8\pi  z)\,\textbf{e}_x$ has the same spatial symmetry as (\ref{spatial-symmetry-exam3}), too.   
Thus, according to the SSIT,  the exact solution must maintain the same spatial symmetry (\ref{spatial-symmetry-exam3}) forever, i.e., for $t\in[0,+\infty)$. 

\begin{figure}
    \begin{center}
        \begin{tabular}{cc}
             \includegraphics[width=2in]{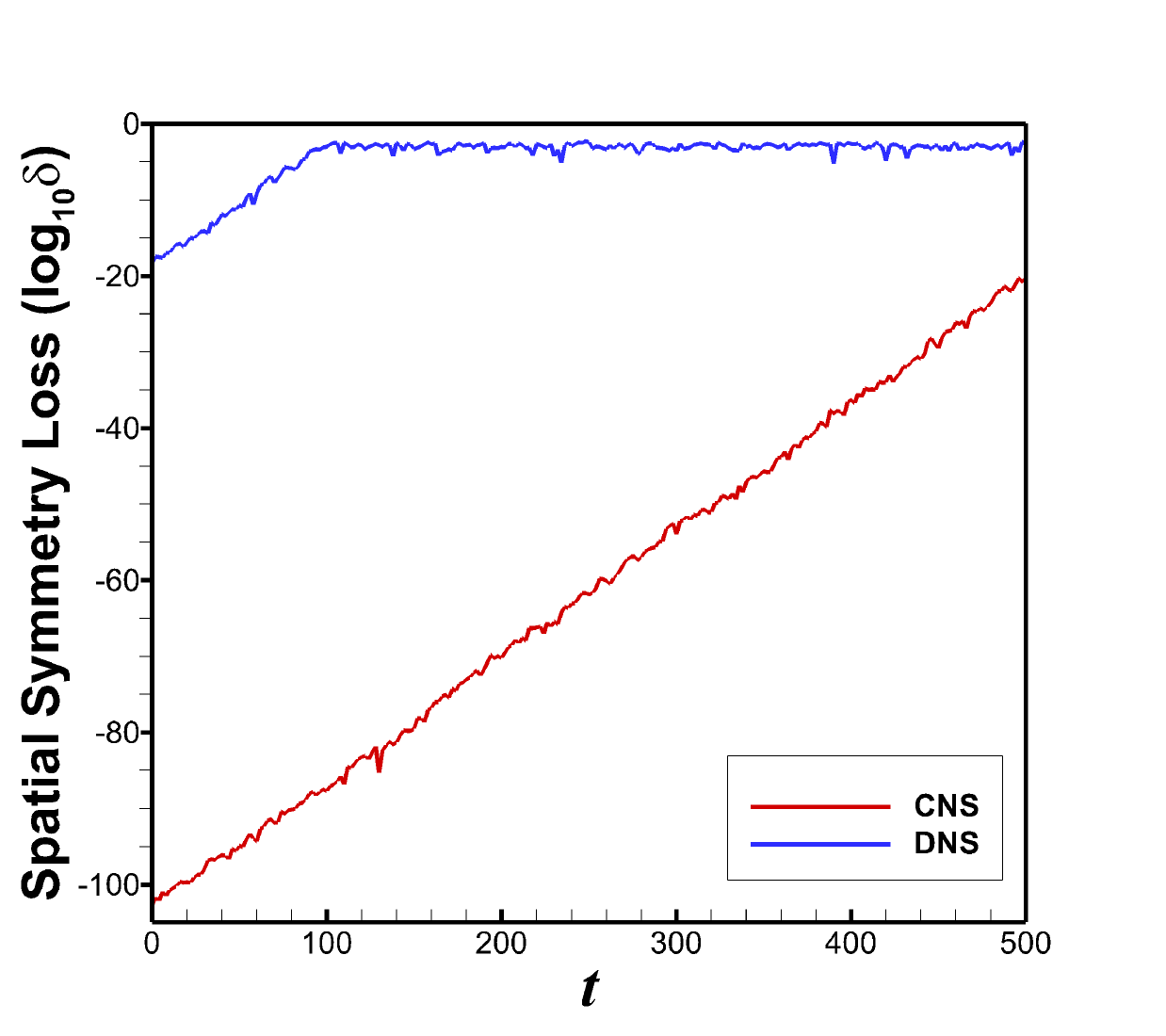}
             \includegraphics[width=2in]{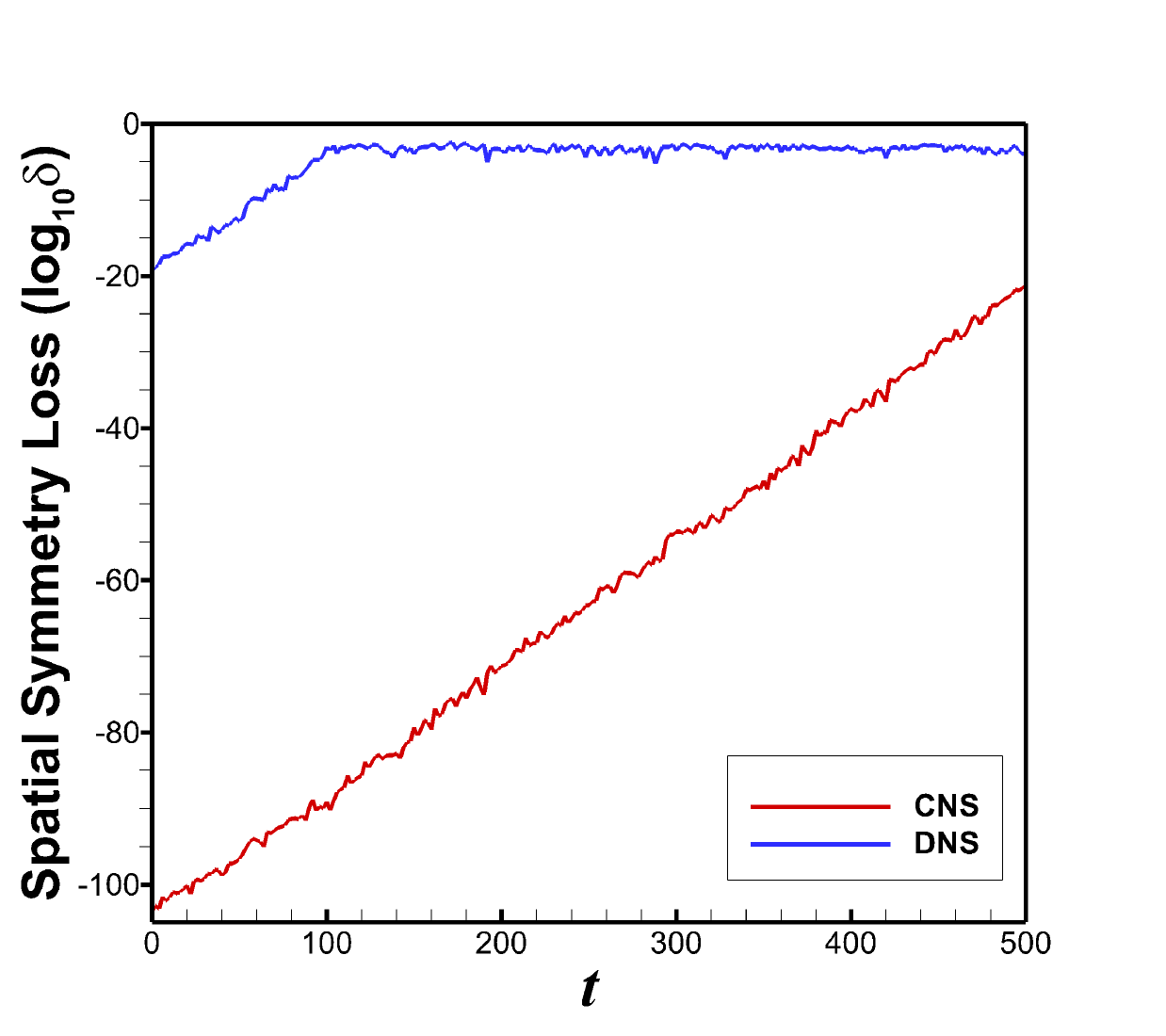}
             \includegraphics[width=2in]{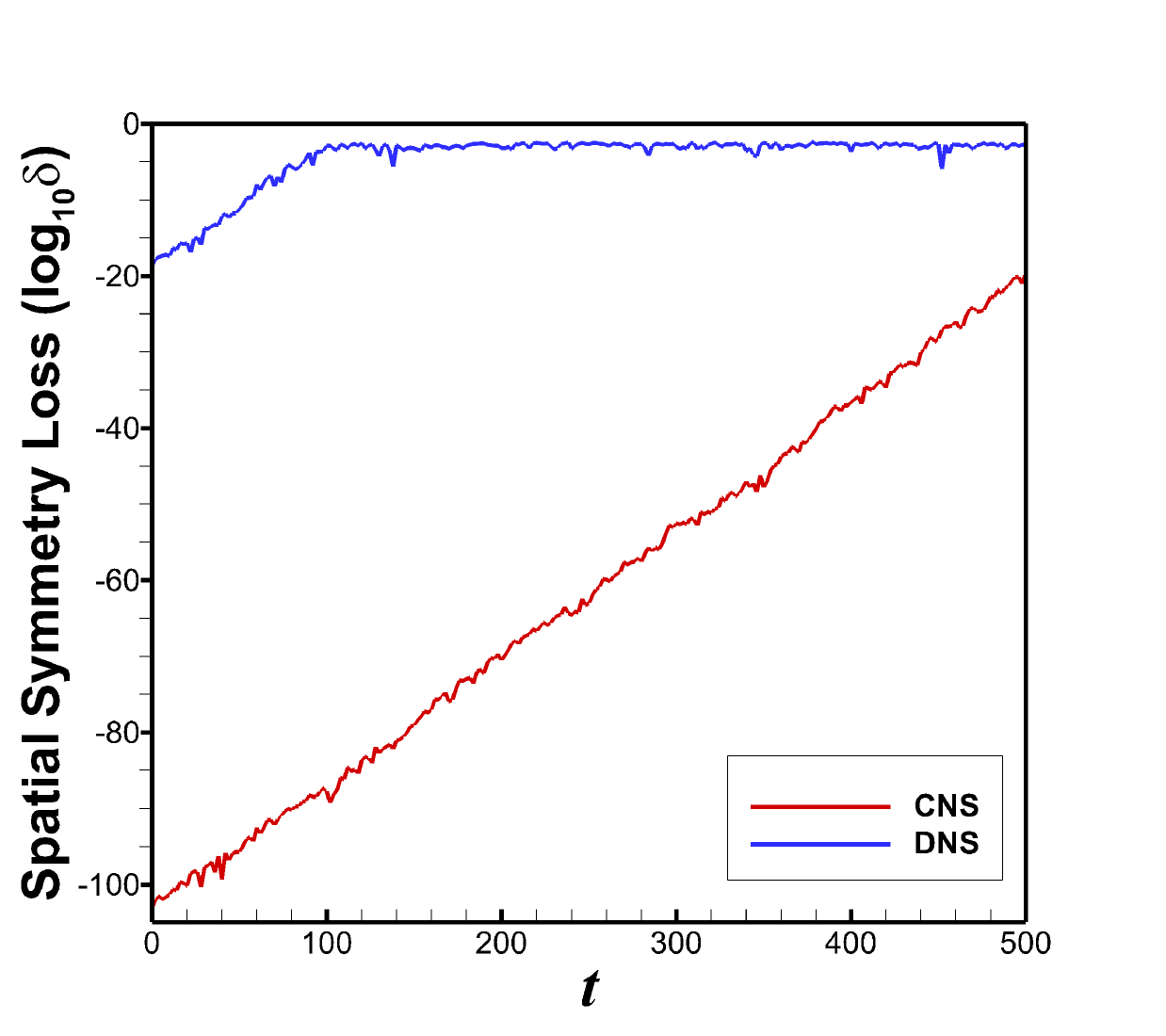}
        \end{tabular}
    \caption{Time histories of  spatial symmetry loss of velocity given by CNS (red) and DNS (blue) for the 3D Kolmogorov flow~\cite{Qin2025PoF}.  Left: $u(x,y,z,t)$; Middle: $v(x,y,z,t)$; Right:  $w(x,y,z,t)$.  The spatial symmetry ($s_{3}$) is defined by (\ref{spatial-symmetry-exam3}).  }     \label{POF-1}
    \end{center}
\end{figure}

Qin and Liao \cite{Qin2025PoF} applied both of CNS and DNS to solve the above case of  3D Kolmogorov flow.  
As shown in Fig.~\ref{POF-1},  the spatial symmetry loss of the CNS result is much less than 1 in the whole simulation of duration $t\in[0,500]$, indicating that the CNS result maintains the same spatial symmetry (\ref{spatial-symmetry-exam3}) as the initial condition (\ref{initial-condition-exam3}) throughout the whole  duration of simulation.  However, as shown in Fig.~\ref{POF-1}, the spatial symmetry loss of the DNS result quickly increases to macro-level $O(1)$ at $t\approx 100$, indicating that the corresponding DNS result maintains the spatial symmetry only over a brief initial period and then completely loses the symmetry when  $t>100$.  
In other words, the CNS result complies with the SSIT throughout  the whole simulation,  but the DNS result  violates the SSIT once $t > 100$.  
Therefore, based on the SSIT, we find that the spatiotemporal trajectories given by DNS are quickly,  badly polluted by numerical noise, which is random in space and destroys the spatial symmetry when  reaching the macro-level.  
This again clearly illustrates that loss of symmetry in traditional DNS is not created by a `physical
instability' but is rather a `numerical artifact' caused by numerical nose and the chaotic nature of NS turbulence.

\subsection{Revealing mathematical reality of NS turbulence}

\subsubsection{Ultra-chaotic property} \label{section:ultrachaos}

\begin{figure}[tb]
    \begin{center}
        \begin{tabular}{cc}
             \includegraphics[width=2in]{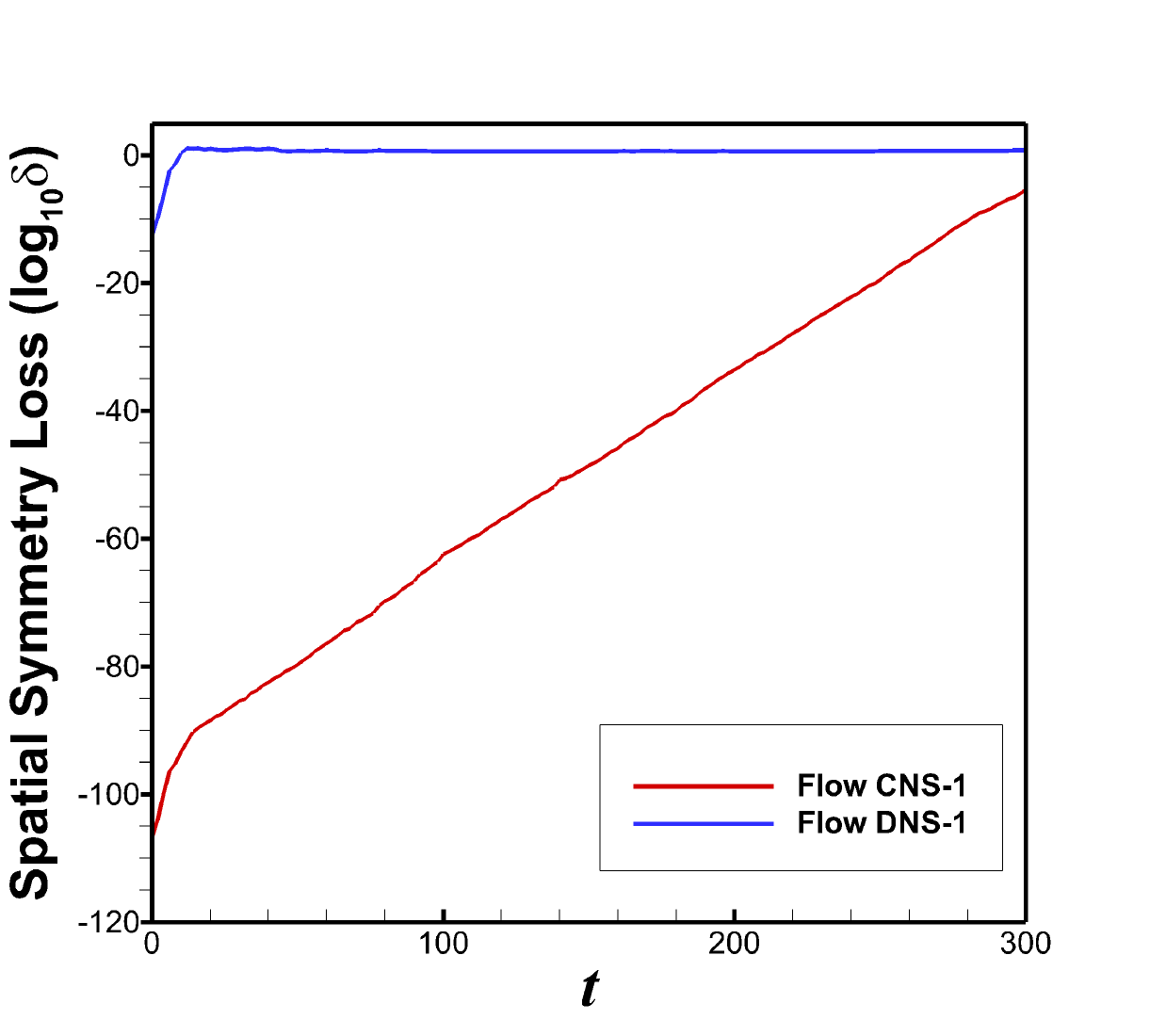}
             \includegraphics[width=2in]{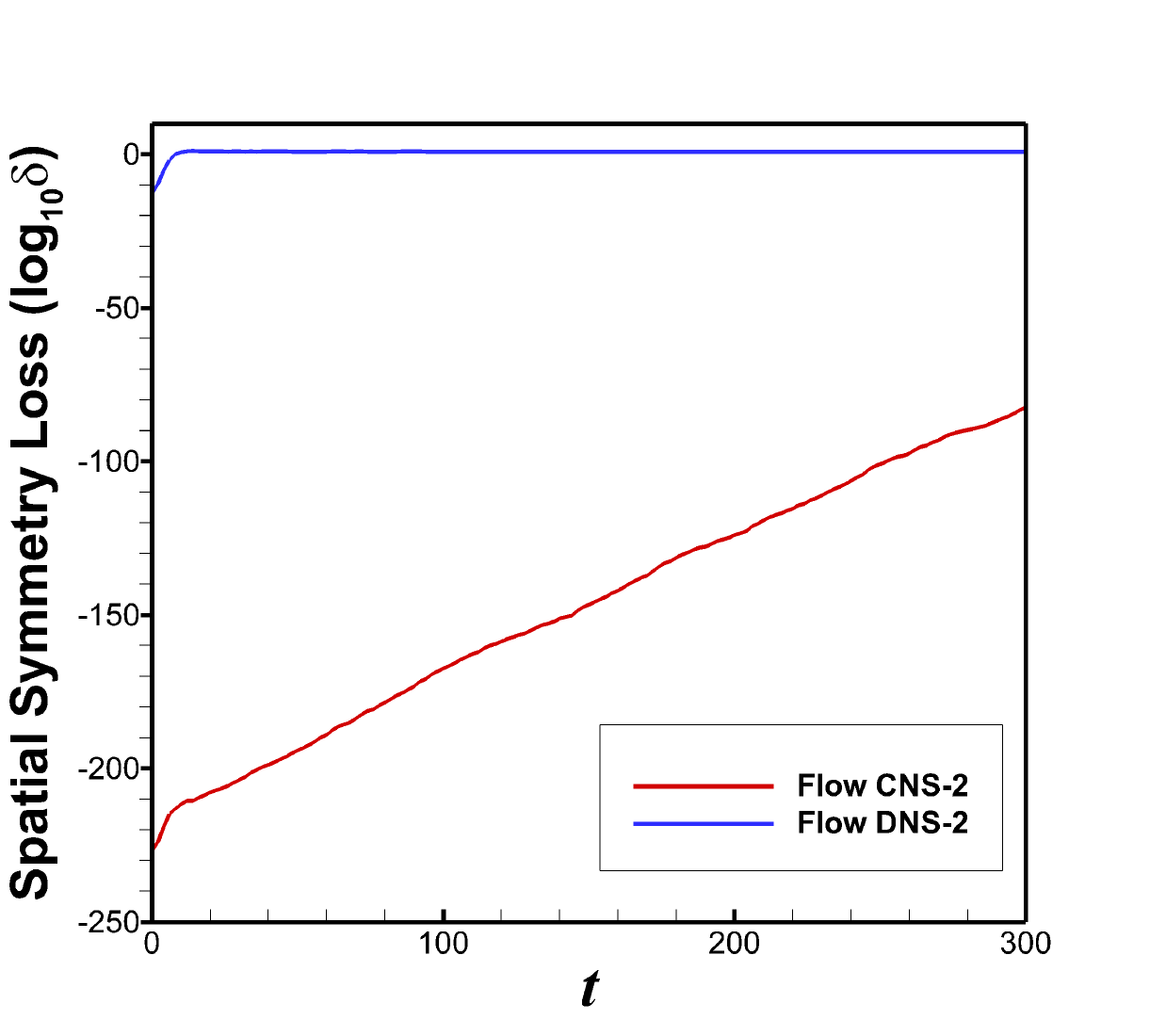}
             \includegraphics[width=2in]{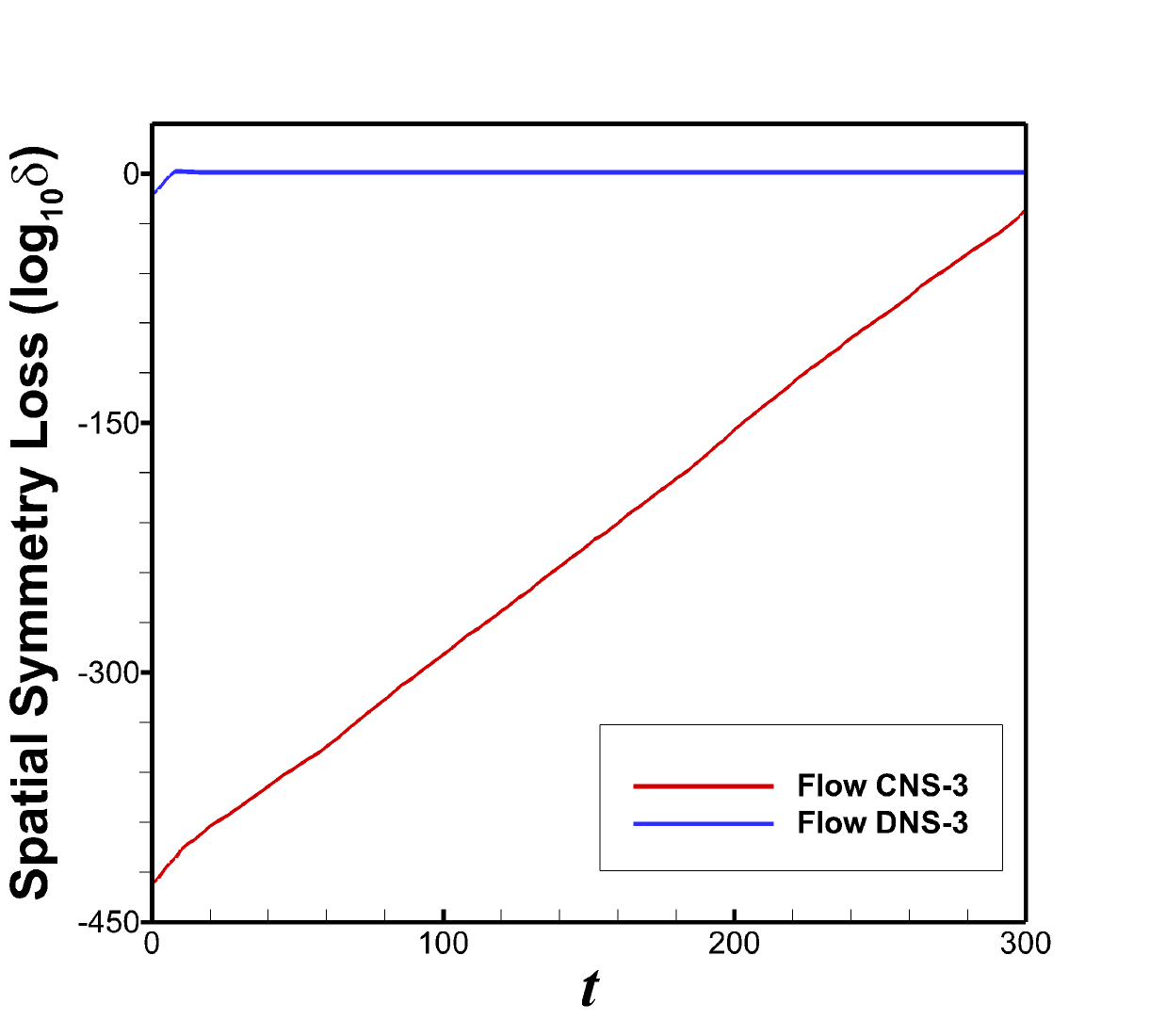}
        \end{tabular}
    \caption{Time histories of spatial symmetry loss given by CNS (red) and DNS (blue) for the 2D Kolmogorov flow~ \cite{Qin2026EJMB}.  Left:  Comparison between Flow~CNS-1 (red) and Flow~DNS-1 (blue) for the spatial symmetry (\ref{symmetry-exam2-1}); Middle:Comparison between  Flow~CNS-2 (red) and Flow~DNS-2 (blue) for the spatial symmetry (\ref{symmetry-exam2-2});  Right: Comparison between Flow~CNS-3 (red) and Flow~DNS-3 (blue) for the spatial symmetry (\ref{symmetry-exam2-3}).}     \label{EJMBF}
    \end{center}
\end{figure}

Let us consider such a set of different initial flows of NS turbulence
\begin{equation}
\textbf{u}_{i} = \textbf{u}_{0} + \epsilon \; \textbf{u}'_{i}, \hspace{1.0cm} i=1,2,\cdots,
\end{equation}
that $\textbf{u}_{i}$ has a different spatial symmetry from each other when the constant $\epsilon$ is at a macroscopic level, i.e.  $\epsilon \sim O(1)$.  According to the SSIT, the exact solution of the NS turbulence given by the initial flow $\textbf{u}_{i}$ should have different spatial symmetries that correspond to different statistics.   Then, let us consider the case when $\epsilon$ is very small, for example, $\epsilon=10^{-10}$ so that the above initial flows are almost the same and $\epsilon \; \textbf{u}'_{i}$ can be regarded as a small disturbance of the initial flow.  However, since NS turbulence has the chaotic property, each small disturbance $\epsilon \; \textbf{u}'_{i}$  certainly increases to macro-level, and then,   according to the SSIT,  their exact solutions of the corresponding NS turbulence  should have different spatial symmetries (and also different statistics).  Recalling the fact that these initial flows are almost the same when $\epsilon=10^{-10}$, their statistics are certainly rather sensitive to the initial flows.  In other words, NS turbulence should have ultra-chaotic property, a novel concept proposed by Liao and Qin~\cite{Liao2022AAMM} in 2022.

The above logical  deduction  has been verified by CNS.     
In 2026, Qin {\em et al.}  \cite{Qin2026EJMB}  applied both of CNS and DNS to solve the 2D Kolmogorov equation (\ref{eq-Kolmogorov-2D}) subject to the periodic boundary condition (\ref{psi-bc-periodic}), but using the physical parameters $n_K=16$, $Re=2000$ and  the following three different initial conditions:
\begin{eqnarray}
 \Phi_{1,0}(x,y)=\psi(x,y,0) & = & -\frac{1}{2}\big[\cos(4x+4y) + \cos(4x-4y) + \sin(4x+4y)\big] \nonumber\\&+& 10^{-10}\big[\sin(x+y) + \sin(x-y)\big],       \label{initial-condition-exam2-1} \\
 \Phi_{2,0}(x,y)= \psi(x,y,0) & = & -\frac{1}{2}\big[\cos(4x+4y) + \cos(4x-4y) + \sin(4x+4y)\big] \nonumber\\&+& 10^{-10}\big[\sin(2x+y) + \sin(2x-y)\big],       \label{initial-condition-exam2-2} \\
  \Phi_{3,0}(x,y)= \psi(x,y,0) & = & -\frac{1}{2}\big[\cos(4x+4y) + \cos(4x-4y) + \sin(4x+4y)\big] \nonumber\\&+& 10^{-10}\big[\sin(4x+y) + \sin(4x-y)\big],       \label{initial-condition-exam2-3}
\end{eqnarray}
corresponding to three turbulent Kolmogorov flows with different spatial symmetries ($s_{4}, s_{5}, s_{6}$) of translation 
\begin{eqnarray}
\Phi_{1,0}(x,y) &:& s_{4}\circ \psi(x,y,t) \mapsto \psi(x+\pi,y+\pi,t),    \label{symmetry-exam2-1} \\
\Phi_{2,0}(x,y) &:& s_{5}\circ \psi(x,y,t) \mapsto \psi(x+\pi/2,y+\pi,t),   \label{symmetry-exam2-2} \\
\Phi_{3,0}(x,y) &:& s_{6}\circ \psi(x, y,t) \mapsto \psi(x+\pi/2, y,t).   \label{symmetry-exam2-3}
\end{eqnarray}
It should be emphasized that the above-mentioned  three spatial symmetries are quite different, although the deviations  between the three initial conditions (\ref{initial-condition-exam2-1}), (\ref{initial-condition-exam2-2})  and (\ref{initial-condition-exam2-3})   are extremely  small, say,
\begin{equation}
\left| \Phi_{i,0}(x,y) - \Phi_{j,0}(x,y) \right|  \leq 4 \times 10^{-10},  \hspace{0.5cm} 1\leq i,j\leq 3, i\neq j, 
\end{equation}
  so that they are {\em almost the same}.

As reported by Qin {\em et al.}  \cite{Qin2026EJMB},  the CNS results maintain the same spatial symmetries  (\ref{symmetry-exam2-1}), (\ref{symmetry-exam2-2}) and (\ref{symmetry-exam2-3}) as the corresponding initial conditions  (\ref{initial-condition-exam2-1}), (\ref{initial-condition-exam2-2})  and (\ref{initial-condition-exam2-3}), respectively,  and thus comply with the SSIT throughout the whole simulation  duration $t\in[0,300]$.  
Fig.~\ref{EJMBF}  confirms that this is indeed true in that the spatial symmetry loss  $\delta(t)$  of each set of CNS results is much less than 1, indicating that Flow~CNS-1 corresponding  to initial condition (\ref{initial-condition-exam2-1}), Flow~CNS-2 corresponding to initial condition (\ref{initial-condition-exam2-2}), and Flow~CNS-3 corresponding to initial condition (\ref{initial-condition-exam2-3}) always maintain their corresponding spatial symmetries  (\ref{symmetry-exam2-1}), (\ref{symmetry-exam2-2}) and (\ref{symmetry-exam2-3}).  However,  as shown in Fig.~\ref{EJMBF}, the spatial symmetry loss of each set of DNS results quickly escalates to the  macro-level at $t \approx 100$, clearly indicating that the DNS results  only maintain the  spatial symmetry over a brief interval of time  $t\in[0,100]$, but thereafter quickly lose spatial symmetry  so that  the DNS results for $t>100$ violate the SSIT!   In other words,  according to the SSIT, it is  absolutely certain that the DNS results for  $t>100$ depart far away from the exact solutions of the turbulent Kolmogorov flow.  

Qin et al.~ \cite{Qin2026EJMB} found  that, although the deviations  between the three initial conditions (\ref{initial-condition-exam2-1}), (\ref{initial-condition-exam2-2})  and (\ref{initial-condition-exam2-3}) are very small,  the statistics of the corresponding CNS results are entirely different.  In other words, the statistical results of the NS turbulence exhibit  sensitivity dependence on the initial conditions.  Thus, the corresponding Kolmogorov flow is  {\em ultra-chaotic}~\cite{Liao2022AAMM}.  Note that ultra-chaos, whose statistics are sensitive to small disturbances, has a higher level of disorder than normal-chaos, whose  statistics are {\em not} sensitive to small disturbances.   A typical example of normal-chaos is given by the famous Lorenz equation, for which  it has been established that although a chaotic numerical trajectory given by a traditional algorithm (using double precision arithmetic) deviates significantly at about $t\approx 35$   from the corresponding CNS result~\cite{Liao2014SCPMA}  that is convergent and reliable in $t\in [0,10000]$, its statistics agree well with those given by the CNS result.  Here  the concept `{\em ultra-chaos}' ~\cite{Liao2022AAMM}  complies with the spirit of SSIT because the three initial conditions (\ref{initial-condition-exam2-1}), (\ref{initial-condition-exam2-2})  and (\ref{initial-condition-exam2-3}), although {\em almost} the same,  have different spatial symmetries which should always persist when the small initial differences evolve to the macro-level due to the chaotic property (i.e. butterfly-effect) of NS turbulence.  Thus, the statistics  of the three sets of CNS results certainly  should be obviously different from each other.   This reveals the following non-uniqueness conjecture for NS turbulence: for almost the same initial conditions, distinct globally smooth solutions of NS turbulent flows may be obtained with different spatial symmetries and statistics.  In other words, the ultra-chaotic characteristic \cite{Qin2026EJMB}  of NS turbulence  is  quite  reasonable and straight-forward regarding both of the spatial symmetry invariance theorem (SSIT) and the chaotic property of NS turbulence!      

\begin{figure}[tb]
    \begin{center}
        \begin{tabular}{cc}
             \includegraphics[width=3.0in]{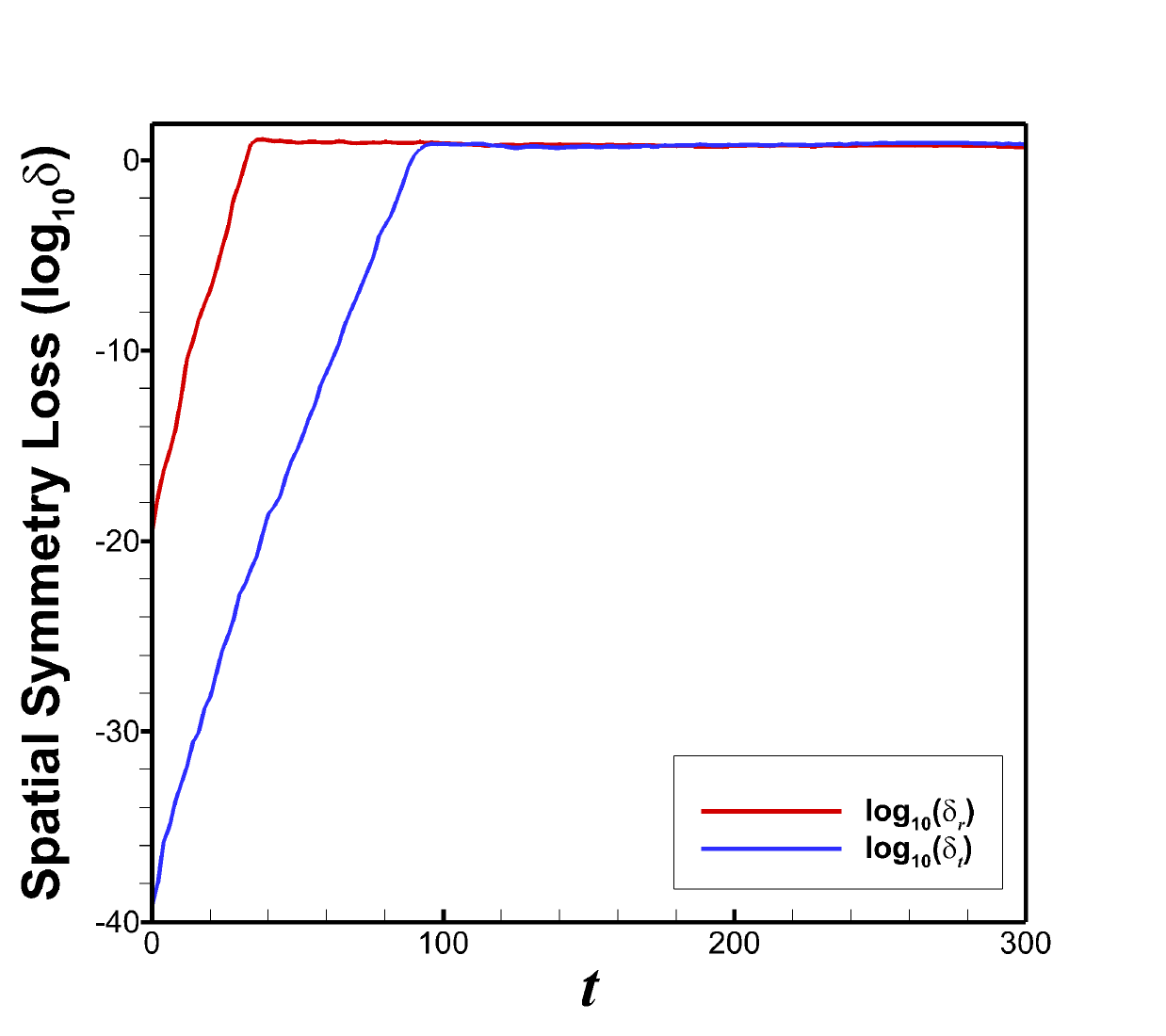}
        \end{tabular}
    \caption{Time histories of spatial symmetry loss of CNS result for the 2D Kolmogorov flow~\cite{ Liao-2025-JFM-NEC} under the initial condition~(\ref{initial-exam4}) when $\delta_{1}=10^{-20}$ and $\delta_{2}=10^{-40}$. Blue line: spatial symmetry of  translation $s_{1}\circ \psi(x,y,t) \mapsto \psi(x+\pi,y+\pi,t)$; Red line:  spatial symmetry of rotation $s_{2}\circ\psi(x,y,t) \mapsto \psi(-x,-y,t)$. }     \label{JFM}
    \end{center}
\end{figure}

\subsubsection{Noise-expansion cascade} \label{section:NEC}
Let us consider such an initial flow of NS turbulence
\begin{equation}
\textbf{u}_{ 0}= \textbf{u}'_{0} + \sum_{i=1}^{2}\epsilon_{i} \; \textbf{u}'_{i}, \hspace{1.0cm} \label{InitialFlow:NEC}
\end{equation}
that $\textbf{u}'_{0}$,  $\textbf{u}'_{0}+\epsilon_{1} \textbf{u}'_{1}$ and  $\textbf{u}'_{0}+\epsilon_{1} \textbf{u}'_{1}+\epsilon_{2} \textbf{u}'_{2}$  have different spatial symmetries, namely   $\alpha, \beta$ and $\gamma$, respectively,  when the constants $\epsilon_{1}$ and $\epsilon_{2}$ are at macroscopic level, i.e.  $\epsilon_{1} \sim O(1)$ and $\epsilon_{2} \sim O(1)$. 
According to the SSIT, the exact solutions of NS turbulence subject to each of these three symmetric initial flows should have different spatial symmetries (and different statistics).  Now let us consider the initial flow (\ref{InitialFlow:NEC}) in the case of $\epsilon_{2}\ll\epsilon_{1}\ll 1$, for example $\epsilon_{1}=10^{-20}$ and $\epsilon_{2}=10^{-40}$.   According to the chaotic property of NS turbulence, the first small disturbance $\epsilon_{1} \textbf{u}'_{1}$ increases to macro-level after a period of time $T_{1}$, and then  the second small disturbance $\epsilon_{2} \textbf{u}'_{2}$ increases to macro-level after a period of time $T_{2}$, respectively.   Thus, according to the SSIT,  the exact solution of the NS turbulence subject to (\ref{InitialFlow:NEC}) should maintain the spatial symmetry $\alpha$ from the beginning until $t \approx T_{1}$ when  the first small disturbance $\epsilon_{1} \textbf{u}'_{1}$ increases to macro-level to replace the spatial symmetry $\alpha$ by $\beta$, and the symmetry $\beta$ should maintain until $t\approx T_{1} + T_{2}$ when   the second small disturbance $\epsilon_{2} \textbf{u}'_{2}$ increases to macro-level to replace the spatial symmetry $\beta$ by $\gamma$.    In other words, NS turbulence should have the so-called `noise-expansion cascade', which can be directly deduced  in logic from the SSIT.

The so-called `noise-expansion cascade' has been first discovered by 
Liao and Qin~\cite{ Liao-2025-JFM-NEC} who  
 applied CNS to solve the 2D Kolmogorov equation (\ref{eq-Kolmogorov-2D}) subject to the periodic boundary condition (\ref{psi-bc-periodic}) for  $n_K=16$ and $Re=2000$, but subject to  the  initial  condition
\begin{equation}
\psi(x,y,0) = -\frac{1}{2}\left[ \cos(x+y)+\cos(x-y)  \right] + \delta_{1} \sin(x+y) +\delta_{2} \sin(2x + y), \label{initial-exam4}
\end{equation}
where $\delta_{1}$ and $\delta_{2}$ are constants.  This involves  the {\em two} kinds of spatial symmetries
\begin{equation}
\left\{ 
\begin{array}{lll}
\mbox{translation} &:& s_{1}\circ \psi(x,y,t) \mapsto \psi(x+\pi,y+\pi,t), \\
\mbox{rotation}  &:& s_{2}\circ\psi(x,y,t) \mapsto \psi(-x,-y,t),
\end{array}
\right.
 \hspace{0.3cm}  \label{symmetry-exam4:1}
\end{equation}
when $\delta_{1}=0, \delta_{2}=0$, and {\em only} the spatial symmetry of rotation
 \begin{equation}
s_{2}\circ\psi(x,y,t) \mapsto \psi(-x,-y,t),  \hspace{0.5cm}\label{symmetry-exam4:2}
\end{equation}
when $\delta_{1}\neq 0, \delta_{2}=0$, but {\em no} spatial symmetry when $\delta_{1}\neq 0, \delta_{2} \neq 0$.  In the special case of $\delta_{1}=10^{-20}$ and $\delta_{2}=10^{-40}$ for the initial condition (\ref{initial-exam4}),  
Liao and Qin~\cite{ Liao-2025-JFM-NEC} found that the CNS results maintain both kinds of spatial symmetry  (\ref{symmetry-exam4:1}) until $t\approx 35$ when the first small disturbance  $\delta_{1}\sin(x+y)$ of the initial flow increases to a macro-level and destroys the rotation symmetry (\ref{symmetry-exam4:1}),  thereafter maintains the  translational symmetry (\ref{symmetry-exam4:2}) until $t\approx 93$ when the second small disturbance  $\delta_{2}\sin(2 x+y)$ of the initial flow increases to macro-level and destroys the spatial symmetry completely.   This is indeed true: as  shown in Fig.~\ref{JFM},  the rotation spatial symmetry loss  increases to macro-level $O(1)$ at $t\approx 35$ with the translation spatial symmetry loss remaining well below.  At $t\approx 93$,  the translation spatial symmetry loss  reaches  macro-level $O(1)$   and the flow loses both spatial symmetries thereafter.    This behaviour is logically reasonable and readily understand from the viewpoint of SSIT.

The results reveal the following mathematical truth underpinning NS turbulence called the `noise-expansion cascade'~\cite{ Liao-2025-JFM-NEC}:   each micro-level noise and/or disturbance of a particular order of magnitude in the initial conditions of NS turbulence grows progressively, following a kind of inverse cascade until reaching the macro level.  Note that, when $\delta_{2} \ll \delta_{1}$, the second disturbance to the initial condition (\ref{initial-exam4}) could be neglected from the traditional DNS viewpoint.   However, this is incorrect, because the DNS results violate the SSIT, as illustrated in~\cite{ Liao-2025-JFM-NEC}.    In other words, the SSIT  inevitably  leads to the `noise-expansion cascade', first discovered by Liao and Qin~\cite{ Liao-2025-JFM-NEC} using CNS.

\subsubsection{Permanent memory of symmetric initial flow condition} \label{section:memory}
According to the SSIT, the exact solution of NS turbulence should {\em remember} the spatial symmetry of initial flow, {\em forever}.  In other words,  the memory about spatial symmetry of initial flow will {\em never} been forgotten by the exact solution of NS turbulence.  Thus, if initial flow has spatial symmetry, the exact solution of the corresponding NS turbulence should have {\em no} mixing.   
This is a very surprising conclusion,  which however is a {\em logical} deduction from the SSIT.

\begin{figure}
    \begin{center}
        \begin{tabular}{cc}
             \includegraphics[width=3.0in]{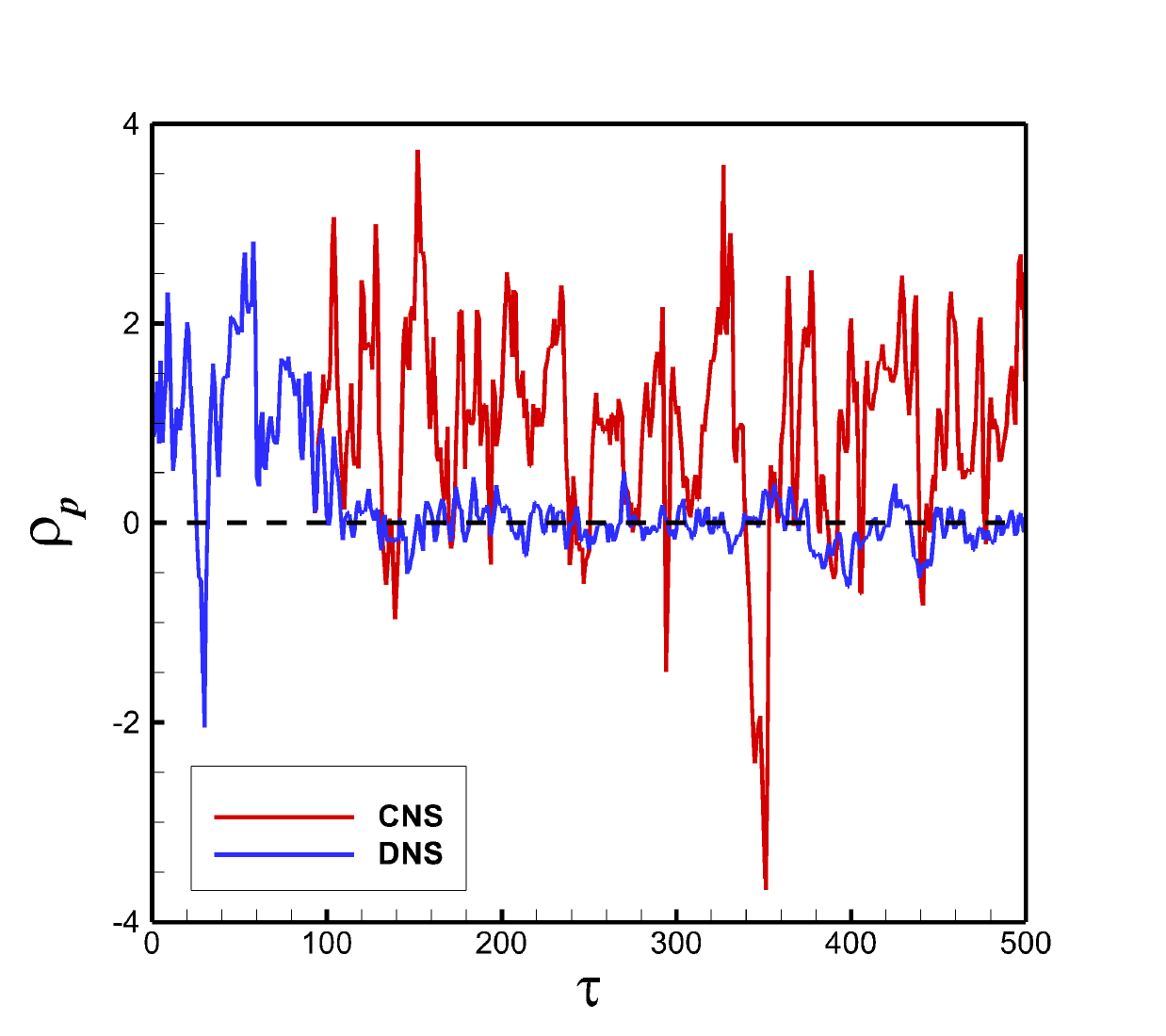}
        \end{tabular}
    \caption{Normalized temporal autocorrelation coefficient $\rho_p(\tau)$ of the pressure field $p(x,y,z,\tau)$ with respect to the initial pressure field $p(x,y,z,0)$ of the 3D Kolmogorov flow~\cite{Qin2025PoF}, given by the CNS benchmark solution (red line) and the DNS result (blue line), where the black dashed line denotes $\rho=0$.}     \label{p-auto}   
    \end{center}
\end{figure}

The above logical  deduction  was confirmed by   Qin and Liao~\cite{Qin2025PoF}  who used both of CNS and DNS to solve a 3D Kolmogorov flow \cite{wu2021quadratic},  governed by the 3D incompressible Navier-Stokes equation.  As proved in \S~2, the exact solution of its pressure field $p(x,y,z,t)$ should remain the spatial symmetry in the whole interval of time $t\in[0,+\infty)$.  
Consider the  temporal autocorrelation coefficient $\rho(\tau)$ of the pressure field $p(x,y,z,\tau)$ with respect to the initial pressure $p(x, y, z, 0)$ normalized by $p(x, y, z,0)$, i.e. 
\begin{equation}
\rho(\tau) = \frac{\big< p(x, y, z,0) \; p(x, y, z,\tau) \big>}{\big<  p^{2}(x, y, z,0) \big>},
\end{equation}  
where $\big< \big>$ is an operator of spatial average.    Note that  $\rho(\tau)$ might be larger than 1 since it is normalized by $p(x, y, z,0)$.  As shown in Fig.~\ref{p-auto},  
the autocorrelation coefficient $\rho(\tau)$ of the CNS solution of the 3D Kolmogorov flow~\cite{Qin2025PoF} is significantly non-zero in the whole interval  $t\in[0,500]$ of simulation, say, almost always departs far away from zero, clearly indicating that  the CNS solution  almost always {\em remembers}  its  initial  condition.  In other words,  the  CNS  results of the 3D Kolmogorov flow~\cite{Qin2025PoF}  have {\em no}  mixing.  
So, the CNS result~\cite{Qin2025PoF} complies with the logical deduction from the SSIT mentioned above, say,  the exact solution of NS turbulence has permanent memory of initial flow.  However, the autocorrelation coefficient $\rho(\tau)$ of the DNS solution of the 3D Kolmogorov flow~\cite{Qin2025PoF}  quickly decreases to be close to zero, indicating that the DNS trajectory, which is quickly polluted by random numerical noise, forgets its initial flow quickly, and thus has `mixing'.  Unfortunately, this kind of `mixing' of DNS  violates the logical deduction from the SSIT mentioned above.   The SSIT and our CNS results~\cite{Qin2025PoF} highly  suggest that  spatial  asymmetry, and/or more importantly, randomness disturbances,  should  be {\em necessary}  to `mixing' of NS turbulence (and even physical turbulence).

\subsubsection{Topological restriction on the evolution}

It should be emphasized that the SSIT imposes a topological straitjacket on the evolution of NS turbulence with symmetric initial flow conditions. This is for two reasons: 
\begin{itemize}
\item  Mandatory synchronization: in a symmetric flow, vortices cannot merge randomly; they must merge in synchronized pairs or clusters to preserve the global topological structure.
\item Path locking: spatial symmetry dictates `allowed' and `forbidden' paths for energy transfer, effectively governing the rate and direction of the energy cascade in NS turbulence.
\end{itemize}
Note that mandatory synchronization and  path locking of exact solution of NS turbulence have been confirmed by the CNS~\cite{Qin2024JOES,Liao-2025-JFM-NEC,Qin2025PoF,Qin2026EJMB}.

It should be emphasized that the above-mentioned interesting features of exact solution of NS turbulence, such as ultra-chaos,  the so-called noise-expansion cascade, permanent memory of symmetric initial flow, and topological restriction on the evolution, can be directly deduced in logic from the SSIT.  The fact that the CNS results~\cite{Qin2024JOES,Liao-2025-JFM-NEC,Qin2025PoF,Qin2026EJMB} comply with the SSIT and all of these interesting features highly suggest  that CNS  could indeed reveal the  mathematical truth concerning the {\em exact} solution of NS turbulence.   Unfortunately, DNS results violate the SSIT and thus cannot reveal these mathematical truth,   clearly indicating that DNS results would rapidly become badly polluted by numerical noise and thus depart far from the exact solution of NS turbulence for given initial conditions.

\section{Discussion}

First, let us consider the viewpoints of Corless~\cite{Corless1994}, winner of the Paul R. Halmos - Lester R. Ford Prize of the Mathematical Association of America, that ``{\em  there
are four levels of abstraction used:
\begin{itemize}
\item the physical reality of the problem under study, 
\item the continuous mathematical model of that physical reality, 
\item the numerical discretization of that mathematical model, 
\item and the floating-point simulation of that discretization.
\end{itemize}
Results from one level may or may not transfer easily to another level, and in particular, even qualitative features may not be preserved in that transfer}.''  The present paper focuses on two aspects:  the continuous mathematical model of physical reality (i.e. the NS turbulence governed by continuity equation and deterministic  NS equations), the numerical discretization of that mathematical model and the floating-point simulation of that discretization (i.e. CNS and traditional DNS of NS turbulence).  
As pointed out by Corless~\cite{Corless1994}, we should note that there may be certain numerical simulations that may not preserve even qualitative features of NS turbulence.

\subsection{Scientific value of SSIT}

Do numerical simulations of NS turbulence preserve features of NS turbulence? To answer this important
question, we should consider key features of NS turbulence. For example, the most widely used feature of NS turbulence relates to the famous Kolmogorov -5/3 power law, which has a statistical form.  
The present paper reaches beyond this by providing a mathematical proof of the spatial symmetry invariance theorem (SSIT), whereby  the spatial symmetry of exact  {\em  spatiotemporal} solution of 3D incompressible  NS turbulence subject to periodic boundary conditions and symmetric external force maintains the same symmetry as its initial flow, {\em forever}.  The SSIT has important, scientific value for two reasons:
\begin{enumerate}
\item[(A)]  SSIT provides an `absolute scale' for CFD: if a DNS fails to maintain spatial symmetry of a symmetric initial state, the result is mathematically invalidated as a true solution of NS turbulence; 

\item[(B)] SSIT provides the theoretical foundation (and necessity) for CNS, demonstrating that ultra-high precision is not just a preference but a requirement in order to capture the `mathematically clean' trajectory of NS turbulence.
\end{enumerate}

Our published CNS results confirm the SSIT by means of several benchmark tests. The detailed comparisons between CNS and DNS results strongly suggest that:
\begin{enumerate}
\item[(a)]  the spatiotemporal trajectory of NS turbulence given by DNS is rapidly badly polluted by numerical noise, given that the DNS result quickly violates the SSIT;

\item[(b)]  loss of spatial symmetry in traditional DNS of NS turbulence is not a `physical instability' but instead a `numerical artifact' caused by unavoidable numerical noise and the chaotic nature of the turbulent flow;

\item[(c)]  the CNS results comply with the SSIT over a prescribed finite period of sufficiently long duration to reveal the mathematical ground-truth of NS turbulence, such as its ultra-chaotic characteristics, the noise-expansion cascade,  the permanent memory of symmetric initial flow conditions, and so on. 
\end{enumerate}

Notably, the NS equations lie at the heart of one of the famous Clay Millennium Problems.
The SSIT could offer strategic insights into the Navier-Stokes Existence and Smoothness problem (Millennium Challenge)~\cite{MillenniumProblem} for the following reasons:
\begin{itemize}
\item[(1)] SSIT provides a constraint on numerical singularities, and could restrict the search for potential singularities to within the symmetric function space;
\item[(2)]  The SSIT is a valuable diagnostic tool. If spatial symmetry is lost before a singularity occurs in a simulation, then the associated numerical breakdown is most likely a numerical artifact rather than a mathematical property of the NS equations;
\item[(3)]  The SSIT leads to the following non-uniqueness conjecture: Due to the ultra-chaotic characteristic of NS turbulence, almost the same initial conditions admit distinct globally smooth solutions of NS equations with different spatial symmetries and statistics.
\end{itemize}

Note that, unlike Kida (1985) who utilized `high-symmetry' as a computational tool, the SSIT represents a fundamental shift in paradigm for the following reasons:
\begin{enumerate}
\item[(I)] Kida focused on computational efficiency and used symmetry to reduce the size of the simulation domain by a factor of 64 to achieve higher Reynolds numbers with limited resources. Kida treated spatial symmetry as an observed utility rather than a proven law;
\item[(II)] The SSIT establishes spatial symmetry as a theoretical benchmark. The focus is on the mathematical certainty that initial spatial symmetry must persist forever, say, in $t\in[0,+\infty)$. Any loss of spatial symmetry in a simulation is redefined not as physical instability, but as numerical pollution - providing mathematical evidence that the simulation has deviated from the true solution of the NS equations.
\end{enumerate}

Expansion of numerical noise is essentially due to the chaotic property (trajectory instability caused by  the  uncontrolled growth of small disturbances)~\cite{Deissler1986PoF, boffetta2001predictability, boffetta2017chaos, berera2018chaotic, Vassilicos2023JFM, Okamoto2025PRF}  of  NS turbulence, but has nothing to do with spatial symmetry itself.   
This is the reason why DNS of laminar flow can maintain spatial symmetry of initial flow, given that laminar flow is insensitive to numerical noise. In other words, the numerical noise of  NS turbulence certainly increases to macro-level no matter whether  or not the initial flow is  symmetric. It is due to the SSIT that we can use spatial symmetry to reveal properly  the exponential expansion of numerical noise in DNS.   Similarly, the `noise-expansion cascade' is also a fundamental property of NS turbulence for general flows.  Thus,  the `noise-expansion cascade' is  a fundamental property of NS turbulence for general initial flows, say, no matter whether  or not  the initial flow is symmetric.  Note that it is the SSIT that provides a mathematical cornerstone for the  ultra-chaotic characteristics, the so-called noise-expansion cascade, and permanent memory of initial flow condition of NS turbulence.

It is well known that a numerical solution of an {\em uniformly hyperbolic} system may depart from the exact trajectory while still shadowing a true trajectory of the same dynamical system (the shadowing lemma)~\cite{ANOSOV1967, BOWEN1975}.   However,  hardly any chaotic system is uniformly hyperbolic, and  the non-existence of such shadowing trajectories may be caused by the finite-time Lyapunov exponent fluctuating about zero~\cite{dawson1994obstructions}.    Besides,  no true trajectory of reasonable length can be found to exist near any computer-generated trajectories  for non-hyperbolic chaotic systems with unstable dimension variability~ \cite{do2004statistics}.   
It seems that the good promise suggested by the Shadow Lemma \cite{ANOSOV1967, BOWEN1975} for a uniformly hyperbolic system does not work for many non-hyperbolic systems \cite{Anishchenko2000}.  Most investigations on shadowing  are  based  on  low dimensional dynamic systems. 
It has been found~\cite{sauer2002shadowing} that, even for a low-dimensional dynamic system with only two degrees of freedom, extremely small levels of numerical noise might generate macroscopic errors in simulation average statistics that are several orders of  magnitude  larger than the noise level.  
To the best of the author's knowledge, practical shadowing algorithms have not been previously proposed
for spatiotemporal chaotic systems such as turbulent flows governed by nonlinear PDEs.  
In short, it is certain that the spatiotemporal trajectory of DNS of NS turbulence subject to a given initial flow  should be quickly badly polluted by numerical noise.

 \subsection{Some open questions}

Note that the mathematical proof of the SSIT is based on the assumption that the solution is analytic with a non-zero Taylor series in $t\in[0,+\infty)$.   For the timescales of interest in most physical processes, this assumption holds and nowadays is the foundation of CFD; thus the SSIT provides them only rigorous bound available.  However, it is an open question whether or not this assumption is absolutely necessary for the proof of the spatial symmetry invariance of NS turbulence.

Ergodicity requires that time averages are the same for almost all initial conditions;  in other words, the flow
should rapidly forget its initial flow conditions; however, this inherently fails if different initial conditions lead to
different attractors (corresponding to different statistics).  Unfortunately, NS turbulence has an ultra-chaotic property,  which can be deduced directly using logic from the SSIT, as mentioned in ~\S~\ref{section:ultrachaos}, meaning that different initial conditions might lead to different attractors (and statistics).
   In addition,  it can be logically deduced from the SSIT that  NS turbulence has {\em permanent} memory of symmetric initial flow condition, as mentioned in~\S~\ref{section:memory}.   Thus, according to the SSIT, the exact solution of NS turbulence subject to spatially symmetric initial flow should have {\em no} ergodicity.  
This is  a very {\em surprising} mathematical conclusion based on the SSIT.

The mathematical proofs of ergodicity given by Kuksin and Shirikyan~\cite{Kuksin2001CMP, Kuksin2002JMPA, Kuksin2012Book} for the 2D NS  equations rely on random forcing to provide the  `mixing'  that  destroys memory of initial conditions.   Random white noise acts as a powerful mixing agent: it continuously injects energy across the spatial modes it acts on and forces the system to explore all possible states.   A deterministic force, even if chaotic, does not provide this uniform stochastic exploration.  Without such randomness, one cannot guarantee that correlations decay to zero.  Thus,  as pointed out by Kuksin and Shirikyan~\cite{Kuksin2001CMP, Kuksin2002JMPA, Kuksin2012Book}, the random forcing is a physically essential ingredient that breaks the system's symmetry, destroys multiple attractors, and enforces a unique, mixing stationary state.  Without it, the system is generally not ergodic, and its long-term statistics depend critically on the initial condition and the exact form of the forcing.   
The permanent memory of NS turbulence concerning symmetric initial flow conditions, which is revealed
herein by the SSIT, complies with  the mathematical proofs of ergodicity given by Kuksin and Shirikyan~\cite{Kuksin2001CMP, Kuksin2002JMPA, Kuksin2012Book} for the 2D NS equations, since it clearly  indicates  the  importance and necessity  of  randomness and/or spatial asymmetry in initial flow  to  ergodicity of NS turbulence, which is a {\bf cornerstone} of turbulence theory.  However, it is still an open question whether or not random disturbance in initial flow and/or external force should be absolutely necessary for three-dimensional NS turbulence to have ergodicity.

 Corless~\cite{Corless1994} pointed out that ``{\em there are four levels of abstraction used: 
 the physical reality of the problem under study, 
 the continuous mathematical model of that physical reality, 
the numerical discretization of that mathematical model, 
 and the floating-point simulation of that discretization.
Results from one level may or {\bf may not} transfer easily to another level, and in particular, even qualitative features may not be preserved in that transfer}.''   
 Note that all CNS results comply with the SSIT over a  prescribed simulation period that is finite but long enough to reveal mathematically exact solutions of NS turbulence, and thus may  preserve  qualitative features of NS turbulence such as the spatial symmetry invariance, ultra-chaotic property, noise-expansion cascade,  and permanent memory of the symmetric initial flow condition of NS turbulence.   Unfortunately,    
 the SSIT is violated by all current state-of-the-art DNS of NS turbulence~\cite{Qin2024JOES, Qin2025PoF, Qin2026EJMB}, indicating  that  DNS  {\bf may  not}  preserve  qualitative features of NS turbulence.  Thus, the SSIT  as a {\bf theoretical benchmark} reveals the non-preservation and violation between numerical simulation given by traditional DNS and these qualitative features of exact solution of NS turbulence.  How to eliminate such kind of violation and non-preservation of traditional DNS for NS turbulence?  This is an open question for future researches.  
  
Besides, as mentioned by Corless~\cite{Corless1994}, the real turbulence (including physical experiments of real turbulent flow), and its mathematical model (i.e.  the NS turbulence here), are two levels of abstraction.  Similarly,  the SSIT as a {\bf theoretical benchmark} of NS turbulence  may be used  to verify whether  the features of {\bf real} turbulent flows may be {\bf preserved} in NS turbulence or not.  In other words, it is an open question whether or not  NS turbulence as a {\bf mathematical model} of  turbulence may preserve {\bf all} features of {\bf real turbulence}.  
Note that this paper focused on NS turbulence governed by the continuity and {\bf deterministic} NS momentum equations, which neglect all small stochastic disturbances, including artificial numerical noise and physical disturbance.   It is also an open questions whether  these small stochastic disturbances  may exert significant influences on real turbulent flows so that it would be better to use {\bf stochastic} NS equations rather than deterministic ones.

\section{Conclusion}      

A full mathematical proof was given of a Spatial Symmetry Invariance Theorem (SSIT) for the incompressible fluid flow equations. The theorem states that the exact solution of the continuity and deterministic NS momentum equations with periodic boundary conditions maintains the same spatial symmetry as that of the symmetric initial flow, {\em forever}.   The SSIT has significant scientific value in the fluid mechanics of turbulent flows for the following reasons:

 \begin{enumerate}
\item[(i)]  
Using the SSIT,  it is straightforward to deduce several key properties of NS turbulence, including ultra-chaos, the so-called noise-expansion cascade,  permanent memory of symmetric initial flow conditions, and so on.  All of them are confirmed and verified by CNS. 

 \item[(ii)]  SSIT provides an absolute scale by which to check the validity and accuracy of numerical algorithms used to simulate NS turbulence.  
 \item[(iii)]   It was found that CNS  complies with SSIT and the  interesting properties deduced from it mentioned in (i).  Thus, CNS indeed could obtain true spatiotemporal trajectories of NS turbulence in a finite but long enough period of time.  CNS provides a reliable numerical tool to  investigate NS turbulence.  
 \item[(iv)]  It was found that  DNS quickly violates SSIT, revealing that the spatiotemporal trajectory of NS turbulence predicted by traditional DNS soon becomes badly polluted by numerical noise. Importantly, it is found that even the statistical results of DNS might be unstable, i.e., exhibit sensitivity dependence on the initial flow conditions. In other words, the loss of spatial symmetry in traditional DNS of NS turbulence is not a physical instability but instead a numerical artifact caused by the chaotic nature of the flow.   This implies that modellers should be very cautious when taking traditional DNS results as benchmark solutions for NS turbulence.
 \item[(v)] SSIT and the interesting properties of NS turbulence mentioned in (i) may be used as a theoretical benchmark to check and verify (by comparison with physical experiments) whether or not deterministic NS equations as a mathematical model of turbulence may indeed preserve all features of real turbulence.    
\end{enumerate}
Thus, SSIT provides a new mathematical cornerstone of NS turbulence.  SSIT also gives us a novel way of reconsidering the foundation of the deterministic NS equations in turbulence modelling.    SSIT is therefore capable of greatly deepening our physical insight concerning turbulence. In the future, SSIT could provide a completely new theoretical cornerstone and a completely novel perspective regarding turbulence, considered by Feynman to be the most important unsolved problem of classical physics.  Some open questions are proposed for future investigations.          
   
\section*{Acknowledgements}
Sincere thanks to the anonymous reviewers for their valuable comments.   This work is partly supported by National Natural Science Foundation of China (No. 12521002). 
	
\section*{Data and materials availability}
All data are available by sending requirement to the corresponding author.

\section*{Competing interests}
The authors declare no competing interests.

\bibliographystyle{elsarticle-num}
\bibliography{KolmogorovE}

\end{document}